\documentclass[superscriptaddress,11pt,nofootinbib]{revtex4}
\usepackage{amssymb,graphicx,amsmath,array,verbatim,setspace}

\newcommand{\beq}{\begin{eqnarray}}
\newcommand{\eeq}{\end{eqnarray}}
\newcommand{\p}{\partial}

\newcommand{\vs}[1]{\vspace{#1 mm}}
\newcommand{\hs}[1]{\hspace{#1 mm}}
\newcommand{\bpm}{\begin{pmatrix}}
\newcommand{\epm}{\end{pmatrix}}
\newcommand{\Z}{\mathbb{Z}}
\newcommand{\R}{\mathbb{R}}
\newcommand{\C}{\mathbb{C}}
\newcommand{\tr}{{\rm Tr}}
\newcommand{\D}{\mathcal D}

\newcommand{\ba}{\left(\begin{array}}
\newcommand{\ea}{\end{array} \right)}

\makeatletter
\@addtoreset{equation}{section}

\makeatother

\usepackage[usenames]{color}

\makeatletter
\renewcommand{\p@subsection}{}
\renewcommand{\p@subsubsection}{}
\makeatother

\begin{document}

\title{All-order Resurgence from Complexified Path Integral \\ \vspace{1mm}
in a Quantum Mechanical System with Integrability \vspace{5mm}}

\author{Toshiaki Fujimori}
\email{toshiaki.fujimori018(at)gmail.com}
\address{Department of Physics, and Research and 
Education Center for Natural Sciences, 
Keio University, 4-1-1 Hiyoshi, Yokohama, Kanagawa 223-8521, Japan}

\author{Syo Kamata}
\email{skamata11phys(at)gmail.com}
\address{National Centre for Nuclear Research, 02-093 Warsaw, Poland}

\author{Tatsuhiro Misumi}
\email{misumi(at)phys.kindai.ac.jp}
\address{Department of Physics, Kindai 
University,  Osaka 577-8502, Japan
}
\address{Department of Physics, and Research and 
Education Center for Natural Sciences, 
Keio University, 4-1-1 Hiyoshi, Yokohama, Kanagawa 223-8521, Japan}

\author{\\Muneto Nitta}
\email{nitta(at)phys-h.keio.ac.jp}
\address{Department of Physics, and Research and 
Education Center for Natural Sciences, 
Keio University, 4-1-1 Hiyoshi, Yokohama, Kanagawa 223-8521, Japan}

\author{Norisuke Sakai}
\email{norisuke.sakai(at)gmail.com}
\address{Department of Physics, and Research and 
Education Center for Natural Sciences, 
Keio University, 4-1-1 Hiyoshi, Yokohama, Kanagawa 223-8521, Japan}
\begin{abstract} \vs{10}
We discuss all-order transseries 
in one of the simplest quantum mechanical systems:
a U(1) symmetric single-degree-of-freedom system 
with a first-order time derivative term. 
Following the procedure of the Lefschetz thimble method, 
we explicitly evaluate the path integral 
for the generating function of the Noether charge
and derive its exact transseries expression. 
Using the conservation law, we find all the complex saddle points of the action, 
which are responsible for 
the non-perturbative effects and the resurgence structure of the model. 
The all-order power-series contributions around each saddle point
are generated from the one-loop determinant 
with the help of the differential equations obeyed 
by the generating function. 
The transseries are constructed by summing up 
the contributions from all the relevant saddle points, 
which we identify by determining the intersection numbers 
between the dual thimbles and the original path integration contour. 
We confirm that the Borel ambiguities of the perturbation series are cancelled 
by the non-perturbative ambiguities 
originating from the discontinuous jumps of the intersection numbers.
The transseries computed in the path-integral formalism agrees with the exact generating function, whose explicit form can be obtained in the operator formalism 
thanks to the integrable nature of the model.
This agreement indicates the non-perturbative completeness of the transseries 
obtained by the semi-classical expansion 
of the path integral based on the Lefschetz thimble method. 
\end{abstract}

\maketitle

\newpage
\begin{spacing}{1.3}
\tableofcontents
\end{spacing}
\newpage 

\section{Introduction}
Path integral formalism is one of the fundamental tools to formulate quantum systems. It is based on integration over infinite-dimensional functional spaces of fields. Although it is a general and intuitive formulation, path integrals can rarely be evaluated exactly. One can use the perturbative expansion to approximate a path integral as a power series of a coupling constant. Although such a perturbation series gives a good approximation when the coupling constant is small, it is usually an asymptotic series with a zero radius of convergence. Hence, the perturbation series truncated at a finite order has limited accuracy, particularly for a large expansion parameter. A possible prescription for such a divergent series is Borel resummation. It applies to asymptotic series with factorially divergent expansion coefficients and gives a closed form for the series if it is Borel summable, i.e. its Borel transform is non-singular on the positive real axis of the Borel plane. However, in many physical systems, perturbation series are non-Borel summable and associated with ambiguities depending on the regularization.   

The remedy for such an ill-defined series is to construct the so-called transseries by appropriately summing up the contributions of saddle points, that is, classical solutions of the action. According to the resurgence theory \cite{Ecalle} (see e.g. \cite{Costin:1999798,Marino:2012zq,Dorigoni:2014hea,Aniceto:2018bis,2014arXiv1405.0356S} for reviews on the application of the resurgence theory to field theories), 
all the ambiguities from the perturbative and non-perturbative sectors cancel in the transseries. 
As in the steepest descent (stationary phase) method for ordinary finite-dimensional integrals, we have to take into account complex saddle points which can be found by analytically continuing the action as a holomorphic functional of the fields (see e.g.  \cite{Cherman:2014ofa,
Behtash:2015zha,Behtash:2015loa,Fujimori:2016ljw,Fujimori:2017oab,Behtash:2017rqj,Fujimori:2017osz,Fujimori:2018kqp} for examples of systems where complex saddle points called ``bions" play important roles). Not all the saddle points are relevant, but a specific subset can contribute. Such a subset can be determined by the Lefschetz thimble method, which states that the relevant saddle points are those with steepest-ascent flows (dual thimbles) intersecting with the original path integration contour (the original configuration space). 
The contribution from each saddle point 
is given by the integral over the associated thimble (steepest descent flows). 
Its ambiguity is related to a Stokes phenomenon, a sudden change in the shape of the (dual) thimble, which occurs when the argument of the coupling constant is varied. 
Although the contribution from each saddle point can be ambiguous due to such a Stokes phenomenon, the transseries constructed through the Lefschetz thimble method is unambiguous thanks to the cancellation mechanism of the ambiguities. Such a resurgence structure enables us to find a well-defined closed form with the correct asymptotic expansion. The procedure for evaluating path integrals based on the Lefschetz thimble method can be summarized as follows:
\begin{enumerate}
\item
The first step is to find saddle points by solving the complexified equations of motion derived
from the classical action analytically continued to the complexified configuration space. 
\item
The second step is to determine the contribution of each saddle point. 
It is defined as the path integral over the Lefschetz thimble associated with each saddle point. 
It is, however, usually impossible to directly evaluate such a path integral. 
Instead, one can evaluate the saddle point contribution 
by applying the Borel resummation to the perturbation series around the saddle point configuration.
\item 
The third step is to identify the relevant saddle points 
by examining the intersection numbers between 
the original configuration space and the dual thimbles. 
The transseries can be constructed 
by summing up the contributions of all the saddle points using the intersection numbers as coefficients. 
\end{enumerate}
The question is whether such a transseries is exact or not. It would not be so difficult to see that the Lefschetz thimble method gives exact results for finite-dimensional integrals \cite{Cherman:2014ofa}. On the other hand, path integrals cannot be explicitly evaluated in almost all cases, and hence there is less chance to check the exactness of the transseries. If there exists a model with the following properties, it can serve as a testing ground for the Lefschetz thimble method:
\begin{enumerate}
\item 
All saddle point solutions can be found by solving the complexified equations of motion. 
\item
Expansion coefficients around each saddle point can be determined to all orders. 
\item
All intersection numbers can be determined. 
\item 
Exact results can be obtained via another method.
\end{enumerate} 

It is natural to imagine that the set of the properties described above implies integrability. 
Resurgence structure of integrable fields theories has been discussed in \cite{Marino:2019fuy,Marino:2019wra,Marino:2019eym,Marino:2019fvu,Marino:2020dgc,Marino:2020ggm,Marino:2021six,Marino:2021dzn,Marino:2022ykm} and 
it has been shown that integrability is a powerful tool for studying resurgence structure. 
In this paper, we focus on the case of quantum mechanics that is exactly solvable due to integrability.  
An integrable quantum mechanical system is 
a model with a finite number of degrees of freedom possessing a maximal set of commuting conserved charges. 
In such a system, Hamiltonian can be written by using the action-angle variables $(\boldsymbol \nu, \boldsymbol \theta)$ as a function depending only on the conserved charge $H = V(\boldsymbol \nu)$. 
The simplest class of such integrable quantum mechanical models is the single variable $U(1)$ symmetric first-order time-derivative system. This model can be viewed as a system of a particle on a 2D plane with a large rotationally invariant potential and a constant magnetic field.\footnote{
This system can be viewed as a dimensional reduction of the non-linear Schr\"odinger system in two dimensions, whose resurgence structure is yet to be elucidated from the viewpoint of the Lefschetz thimble method.}
Compared to quantum mechanics with quadratic kinetic terms, 
where resurgence has been extensively discussed \cite{ZinnJustin:2004ib, ZinnJustin:2004cg, Jentschura:2010zza, Jentschura:2011zza, Jentschura:2004jg, Dunne:2013ada,Basar:2013eka,Dunne:2014bca,Misumi:2015dua,Gahramanov:2015yxk,Dunne:2016qix,
Behtash:2015zha,Behtash:2015loa,Fujimori:2016ljw, Sulejmanpasic:2016fwr,Dunne:2016jsr,Kozcaz:2016wvy,Serone:2016qog,Basar:2017hpr,Fujimori:2017oab,Serone:2017nmd,Behtash:2017rqj,Alvarez:2017sza,Fujimori:2017osz,  Behtash:2018voa, Fujimori:2018kqp,Pazarbasi:2019web, Sueishi:2020rug}, 
the first-order time derivative system has half the degrees of freedom and hence a single variable system is 
integrable if there is a conserved charge. 
Therefore, it provides a good playground where we can test the completeness of the Lefschetz thimble method. 
Another important property, which enables us to evaluate the perturbation series around each saddle point, is that the generating function $Z$ for the conserved charge obeys a partial differential equation of the form
\beq
\frac{\p}{\p g} Z = X(g,\p_\mu) Z, 
\label{eq:differential_equation}
\eeq
where $g$ is the coupling constant (expansion parameter), $\mu$ is the external source (imaginary chemical potential) for the conserved charge and $X(g,\p_\mu)$ is a differential operator which depends on the Hamiltonian of the system.
By using the power series ansatz 
on top of the saddle point value $e^{-S_{\rm saddle}/g^2}$,
the differential equation \eqref{eq:differential_equation} 
can be rewritten into a recursion relation for the 
expansion coefficients. 
Starting from the initial term corresponding to  
the one-loop determinant around the saddle point, 
we can solve the recursion relation 
and determine the all-order power series around 
each saddle point. 
Another convenient property of our model is that the intersection numbers are 
accessible in a simple way. 
In particular, we will explicitly determine 
the intersection numbers by solving the gradient flow equation. 
Although the gradient flow equation is originally defined in the complexified configuration space, it is reduced to a finite-dimensional problem using a symmetry argument.
Using these special properties, 
we will show that the transseries obtained 
in the path integral formalism agrees with 
the exact partition function obtained 
in the operator formalism.  
  
The organization of this paper is as follows. 
In section \ref{sec:4th}, 
we discuss the resurgence structure in the first-order time derivative system with a $U(1)$ symmetric quartic potential.
After defining the generating function for the conserved charge $Z(g)$ in section \ref{sec:4th}, we discuss the perturbation series for $Z(g)$ in section \ref{subsec:pert_4th}. All the coefficients of $Z(g)$ are determined by perturbatively solving the differential equation for $Z(g)$. We see that the perturbation series is non-Borel summable due to some singularities of its Borel transform. In section \ref{subsec:complex_saddle}, we calculate the contributions of complex saddle point solutions 
and determine the relevant saddle points by examining the intersection numbers based on the Lefschetz thimble method in section \ref{subsec:intersection}. We see that the ambiguities of the saddle point contributions cancel those of the perturbative part. In section \ref{subsec:canonical}, we compare the generating function obtained in the path integral formalism with 
that calculated in the operator formalism. 
In secion \ref{sec:generalization}, we discuss the generalization to the case of generic $U(1)$ symmetric potential. 
Section \ref{sec:general} outlines a generalization to more general integrable quantum mechanical systems. 
Section \ref{sec:conclusion} is devoted to conclusions and discussion. 
Appendix \ref{appendix:Lefschetz} is a brief review of the Lefschetz thimble method, and Appendix \ref{appendix:diff_eq} is a supplement on the properties of the differential equation for the generating function.

\section{First-order System with a $U(1)$ Symmetric Quartic Potential} \label{sec:4th}
In this section, we discuss the resurgence structure of
the first order time derivative system 
with a $U(1)$ symmetric quartic potential. 
This quantum mechanical system is 
one of the simplest example of the models 
in which transseries for some quantities 
such as partition function can be exactly obtained 
in the path integral formalism.

\subsection{Action, Hamiltonian and Generating Function}
\label{subsec:setup}
Let us consider the 1d system described by the action
\beq
S = \int dt \, L =\int dt \Big[ i \bar \phi \p_t \phi - \frac{g}{2} |\phi|^4 \Big],
\label{eq:Lagrangian}
\eeq
where $\phi$ stands for a complex scalar degree of freedom 
and $g$ is a coupling constant. 
This model can be viewed as a system of a particle on the $(x,y)$-plane $(\phi \propto \sqrt{B}(x+iy))$ with a large magnetic field $B$ and a potential $V \propto B^2 |x+iy|^4$.  
Since the Lagrangian $L$ is linear in the time derivative, 
the canonical conjugate of $\phi$ is 
identified with its complex conjugate $i \bar \phi$. 
The Hamiltonian of this system is given by 
\beq
H = i \bar \phi \p_t \phi - L = \frac{g}{2} |\phi|^4. 
\eeq 
This is the conserved quantity
corresponding to the time translation invariance. 
Another conserved quantity is the Noether charge 
for the phase rotation symmetry $\phi \rightarrow e^{i \alpha} \phi$
\beq
{\mathcal N} = |\phi|^2. 
\eeq
In this section, 
we discuss the resurgence structure of this model 
by investigating the weak coupling expansion of 
the generating function 
for the expectation value of ${\mathcal N}$
\beq
Z = \tr \left[ e^{-\beta (\hat{H} + i \mu \hat{\mathcal N})} \right], 
\label{eq:gf}
\eeq
where $\mu$ is the external source for $\hat{\mathcal N}$ 
and can be interpreted as an imaginary chemical potential\footnote{
The chemical potential $\mu$ can also be viewed as a constant background gauge field (holonomy) $A_0$ for the $U(1)$ symmetry, and hence it has periodicity $\mu \sim \mu + 2\pi/\beta$.}. 
Since the canonical commutation relation in this system is given by\footnote{Throughout this paper, the operator corresponding to the classical variable $\bar \phi$ is denoted by $\hat \phi^\dagger$.}
\beq
[\hat \phi, \hat \phi^\dagger] = 1, 
\eeq
the operators $\hat \phi$ and $\hat \phi^\dagger$ 
do not commute with each other 
and hence we must specify the order of the operators
to define the conserved charges. 
In this paper, we adopt the following ordering 
for the conserved charges
\beq
\hat{\mathcal N} = \hat \phi^\dagger \hat{\phi}, \hs{10} 
\hat H = \frac{g}{2} (\hat \phi^\dagger \hat{\phi})^2.
\eeq
With this convention, we can show that the generating function \eqref{eq:gf} satisfies the ``heat equation"
\beq
\left[ \frac{\p}{\p g} - \frac{1}{2 \beta} \frac{\p^2}{\p \mu^2} \right] Z \ = \ \tr \left[ - \frac{\beta}{2} \left\{ (\hat \phi^\dagger \hat \phi)^2 - \hat{\mathcal N}^2 \right\} e^{-\beta (\hat{H} + i \mu \hat{\mathcal N})} \right] \ = \ 0.
\label{eq:diff_eq0}
\eeq
As we will see, this differential equation enables us to
determine the perturbation series 
to all orders in the coupling constant $g$.  

In the operator formalism, 
the generating function $Z$ can be determined 
by using the number eigenstates. 
The Hamiltonian can be rewritten in terms of 
the number operator $\hat{\mathcal N}$ as
\beq
\hat H = \frac{g}{2} \hat{\mathcal N}^2. 
\eeq
This implies that the energy eigenstates 
are the number eigenstates
\beq
| n \rangle = \frac{1}{\sqrt{n!}} (\hat{\phi}^\dagger)^n | 0 \rangle, \hs{10} \hat \phi | 0 \rangle = 0, \hs{10} \hat H | n \rangle = \frac{g}{2} n^2  | n \rangle.
\eeq
Therefore, the generating function can be written as
\beq
Z \ = \ \tr \left[ e^{-\beta(\hat H + i \mu \hat{\mathcal N)}} \right] \ = \ \sum_{n=0}^\infty \exp \left[ - \frac{\beta g}{2} n^2 - i \beta \mu n \right]. 
\label{eq:Z_canonical}
\eeq
We can confirm that this satisfies the differential equation \eqref{eq:diff_eq0}. In the next section, we calculate the same quantity by applying the Lefschetz thimble method (see Appendix \ref{appendix:Lefschetz} for a briefly review of the Lefschetz thimble method) in the path integral formalism and check that the nontrivial resurgence structure obtained through the Lefschetz thimble method 
leads to the exact result \eqref{eq:Z_canonical} with no ambiguity.

~\paragraph{Generating Function in Path Integral Formalism \\}
Let us consider the generating function $Z$
from the viewpoint of the path integral formalism.
Using the Weyl ordering, 
we can rewrite the operator 
$\hat H + i \mu \hat{\mathcal N}$ as
\beq
\hat H + i \mu \hat{\mathcal N} \ = \ \left[ \frac{g}{2} (|\phi|^4)_{\rm W} + i \mu (|\phi|^2)_{\rm W} \right] + \left[ - \frac{g}{2} (|\phi|^2)_{\rm W} - \frac{i\mu}{2} \right],
\eeq
where $(|\phi|^{2n})_{\rm W}$ denotes the Weyl ordered operator 
\beq
 (|\phi|^{2n})_{\rm W} = \frac{1}{(2n)!} \frac{\p^n}{\p a^n} \frac{\p^n}{\p b^n} (a \hat \phi + b \hat{\phi}^\dagger)^{2n}.
 \label{eq:Weyl}
\eeq
Therefore, in the path integral formalism, 
the generating function is given by
\beq
Z = \int \D \phi \,  \exp \left( - S_E - S_W \right), 
\label{eq:path_integral}
\eeq
where $S_E$ is the (Wick rotated $t \rightarrow - i \tau$) action with the source term and 
$S_W$ is the part generated when the Hamiltonian is rewritten in terms of the Weyl ordered operators
\beq
S_E = \int_0^\beta d\tau \left[ \bar \phi \p_\tau \phi + \frac{g}{2} |\phi|^4 + i \mu |\phi|^2 \right], \hs{10}
S_W = \int_0^\beta d\tau \left( -\frac{g}{2} |\phi|^2 - \frac{i \mu}{2} \right).
\label{eq:Euclidean_action}
\eeq
Corresponding to the trace in Eq.\,\eqref{eq:gf}, 
the path integral should be carried out over 
the configurations satisfying the periodic boundary condition 
\beq
\phi(\tau + \beta) = \phi(\tau). 
\eeq
It is convenient to rescale the variable as
\beq
\phi = \frac{\varphi}{\sqrt{g}} \hs{10}
\tilde \phi = \frac{\tilde \varphi}{\sqrt{g}}. 
\label{eq:rescaling}
\eeq
Then, $S_E$ and $S_W$ become
\beq 
S_E = \frac{1}{g} \int_0^\beta d\tau \left( \bar \varphi \p_\tau \varphi + \frac{1}{2} |\varphi|^4 + i \mu |\varphi|^2 \right), \hs{10}
S_W = \int_0^\beta d\tau \left( -\frac{1}{2} |\varphi|^2 - \frac{i \mu}{2} \right).
\label{eq:S0S1}
\eeq
Thus, identifying the coupling constant $g$ as the Planck constant, we regard $S_E$ and $S_W$ 
as a ``classical action" and an ``operator insertion", respectively. 

\subsection{Perturbation Series}\label{subsec:pert_4th}
Let us first evaluate the path integral 
for the generating function $Z$ in Eq.\,\eqref{eq:path_integral}
by using the perturbative expansion. 
Although the standard diagrammatic 
perturbative expansion is possible, 
we can obtain the perturbation series to all orders 
in the coupling constant $g$ more easily
by using the differential equation 
for the generating function $Z$:
\beq
\left[  \frac{\p}{\p g} - \frac{1}{2\beta} \frac{\p^2}{\p \mu^2} \right] Z \ = \ 0. 
\label{eq:diff_eq}
\eeq
This equation implies that the perturbation series can be obtained 
from the generating function of the free theory 
$Z_0 = Z |_{g=0}$ as
\beq
Z_{\rm pert} \ = \ \sum_{n=0}^\infty \frac{g^n}{n!} \frac{\p^n Z}{\p g^n} \bigg|_{g=0} \ = \ \sum_{n=0}^\infty \frac{1}{n!} \left( \frac{g}{2\beta} \frac{\p^2}{\p \mu^2} \right)^n Z_0.
\label{eq:P.E.}
\eeq
To compute the generating function of the free theory $Z_0$, 
let us use the Fourier series expansion of the original variable 
\beq
\phi = \sum_{p=-\infty}^\infty c_p \, \exp \left( \frac{2\pi i p \tau}{\beta} \right).
\eeq
In terms of the Fourier coefficients $c_p$, 
the free theory action can be rewritten as
\beq
\lim_{g \rightarrow 0} S_E = i \beta \sum_{p=-\infty}^\infty \omega_p |c_p|^2, \hs{10}
\lim_{g \rightarrow 0} S_W = - \frac{i\beta \mu}{2}, 
\eeq
where we have defined 
\beq
\omega_p = \mu + \frac{2\pi p}{\beta} \hs{5} (p \in \Z). 
\eeq
Performing the Gaussian integral for each mode, 
we obtain
\beq
Z_0 \, = \, N \, e^{\frac{i\beta \mu}{2}} \prod_{p=-\infty}^\infty \int d c_p d \bar c_p \, e^{-i \beta \omega_p |c_p|^2} \ = \ N \, e^{\frac{i\beta \mu}{2}} \prod_{p=-\infty}^\infty \frac{\pi}{i(\beta \mu + 2\pi p)} \ = \ \frac{1}{1-e^{-i\beta\mu}},
\label{eq:Z_0_infinite_product}
\eeq
where we have chosen the normalization factor as 
\beq
N = \frac{1}{\pi} \prod_{p=1}^\infty (4 p^2), 
\label{eq:normalization}
\eeq
and used the formula for the infinite product
\beq
\prod_{p=1}^\infty \frac{1}{1-z^2/p^2} = \frac{\pi z}{\sin \pi z}. 
\eeq
One can easily show that 
this choice of the normalization and sign\footnote{
Note that the sign of the infinite product is ambiguous. 
Relabeling $p=p'+q$ with an arbitrary integer $q$,
we find that the sign depends on the choice of $q$
\beq
Z_0 = N \, e^{\frac{i\beta \mu}{2}} \prod_{p=-\infty}^{\infty} \frac{\pi}{i(\beta \mu + 2\pi p)} = N \, e^{\frac{i\beta \mu}{2}} \prod_{p'=-\infty}^{\infty} \frac{\pi}{i((\beta \mu + 2 \pi q)+ 2\pi p')} = \frac{(-1)^q}{1-e^{-i\beta \mu}}. \notag
\eeq
This ambiguity is related to the ``anomaly" of the periodicity (large gauge transformation) $\mu \sim \mu+2\pi/\beta$, which is canceled if $S_W$ is appropriately taken into account.
} 
is consistent with the canonical quantization (see Sec.\,\ref{subsec:canonical}). 
Plugging $Z_0$ into Eq.\,\eqref{eq:P.E.},
we obtain the perturbation series 
\beq
Z_{\rm pert} \ = \ \sum_{n=0}^\infty \frac{1}{n!} \left( \frac{g}{2\beta}  \frac{\p^2}{\p \mu^2} \right)^n \frac{1}{1-e^{-i\beta\mu}}. 
\label{eq:P.S.}
\eeq
This perturbation series is a divergent asymptotic series. 
To show this, let us expand the generating function of the free theory as\footnote{
Here, the summation should be interpreted as
\beq
\sum_{p=-\infty}^\infty \frac{1}{\beta \omega_p} = \frac{1}{\beta \mu} + \sum_{p=1}^{\infty} \left( \frac{1}{\beta \omega_p} + \frac{1}{\beta \omega_{-p}} \right) = \frac{1}{\beta \mu} +\sum_{p=1}^{\infty} \frac{2\beta \mu}{(\beta \mu)^2 - (2\pi p)^2}. \notag 
\eeq
}
\beq
Z_0 \ = \ \frac{1}{1-e^{-i \beta \mu}} \ = \ \frac{1}{2} - i  \sum_{p=-\infty}^\infty \frac{1}{\beta \omega_p}.
\label{eq:free_expanded}
\eeq
Using this expanded form of $Z_0$,
we can rewrite the perturbation series in Eq.\,\eqref{eq:P.S.} as
\beq
Z_{\rm pert}
\ = \ \frac{1}{2} - i \sum_{n=0}^\infty \frac{1}{n!} \left( \frac{g}{2 \beta} \frac{\p^2}{\p \mu^2} \right)^n \sum_{p=-\infty}^{\infty} \frac{1}{\beta \omega_p} 
\ = \ \frac{1}{2} - \sum_{p=-\infty}^{\infty} \frac{i}{\beta \omega_p} \sum_{n=0}^\infty \frac{(2n)!}{n!} \left( \frac{g}{2 \beta \omega_p^2} \right)^n.
\eeq
Since $(2n)!/n! = 4^n \Gamma(n+1/2)/\Gamma(1/2)$, 
this perturbation series is factorially divergent. 
Rewriting the series as 
\beq
\sum_{n=0}^\infty \frac{(2n)!}{n!} x^n = \int_0^\infty dt \, e^{-t} \sum_{n=0}^\infty \frac{(2n)!}{(n!)^2} (x t)^n = \int_0^\infty dt \frac{e^{-t}}{\sqrt{1-4xt}},
\eeq
we obtain the formal Borel resummation of the perturbation series
\beq
Z_{\rm pert} = \int_0^\infty dt \, e^{-t} \left[ \frac{1}{2} -\sum_{p=-\infty}^{\infty} \frac{i}{\beta \omega_p} \frac{1}{\sqrt{1- \frac{2g t}{\beta \omega_p^2}}} \right].
\label{eq:pert_Borel}
\eeq
\begin{figure}[h!]
\centering
\fbox{
\includegraphics[width=100mm, bb = 100 193 700 413]{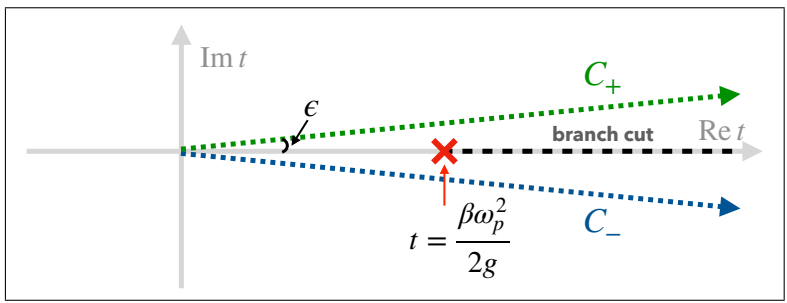}}
\vs{5}
\caption{Contours on Borel plane}
\label{fig:Borel_plane}
\end{figure}
Since there are singularities at 
$t = \beta \omega_p^2/(2g)$, 
we have to regularize the integral to obtain a finite value.  
Giving a small imaginary part to the coupling constant $g$~($\arg g = \epsilon$), or equivalently, performing the Borel resummation along the contours $C_\pm$ shown in Fig.\,\ref{fig:Borel_plane}, 
we can avoid the singularity and obtain a finite value. 
However, the Borel resummation gives different answers
depending on the sign of ${\rm Im} \sqrt{\beta/2g} \,\omega_p$
\beq
Z_{\rm pert} = \frac{1}{2} - \sqrt{\frac{\pi}{2\beta g}} \sum_{p=-\infty}^{\infty} e^{- \frac{\beta}{2g} \omega_p^2} \left[ {\rm erf} \left( i \sqrt{\frac{\beta}{2g}} \omega_p \right) + {\rm sign} \left( {\rm Im} \sqrt{\frac{\beta}{2g}} \omega_p \right) \right], 
\label{eq:Z_pert}
\eeq
where ${\rm erf}(z)$ is the error function defined by
\beq
{\rm erf}(z) = \frac{2}{\sqrt{\pi}} \int_0^z dx \, \exp(-x^2).
\label{eq:error}
\eeq
The discontinuity at ${\rm arg} g = 0$ is given by
\beq
 Z_{\rm pert}^{(+)} - Z_{\rm pert}^{(-)} \ = \ \sqrt{\frac{2\pi}{\beta g}} \sum_{p=-\infty}^{\infty} {\rm sign} (\omega_p) \exp \left(- \frac{\beta \omega_p^2}{2g} \right),
 \label{eq:discon}
\eeq
where $Z_{\rm pert}^{(\pm)}$ are the perturbation series 
for $\arg g > 0$ and $\arg g < 0$, respectively. 
This discontinuity has non-perturbative factors, 
and hence it is expected to be related 
to non-perturbative effects. 
In the next section, we show that there are complex saddle points of the Euclidean action whose non-perturbative contributions
cancel these ambiguities of the perturbation series. 

\subsection{Complex Saddle Points}\label{subsec:complex_saddle}
\begin{figure}[t]
\fbox{
\includegraphics[width=60mm, bb = 30 30 510 510]{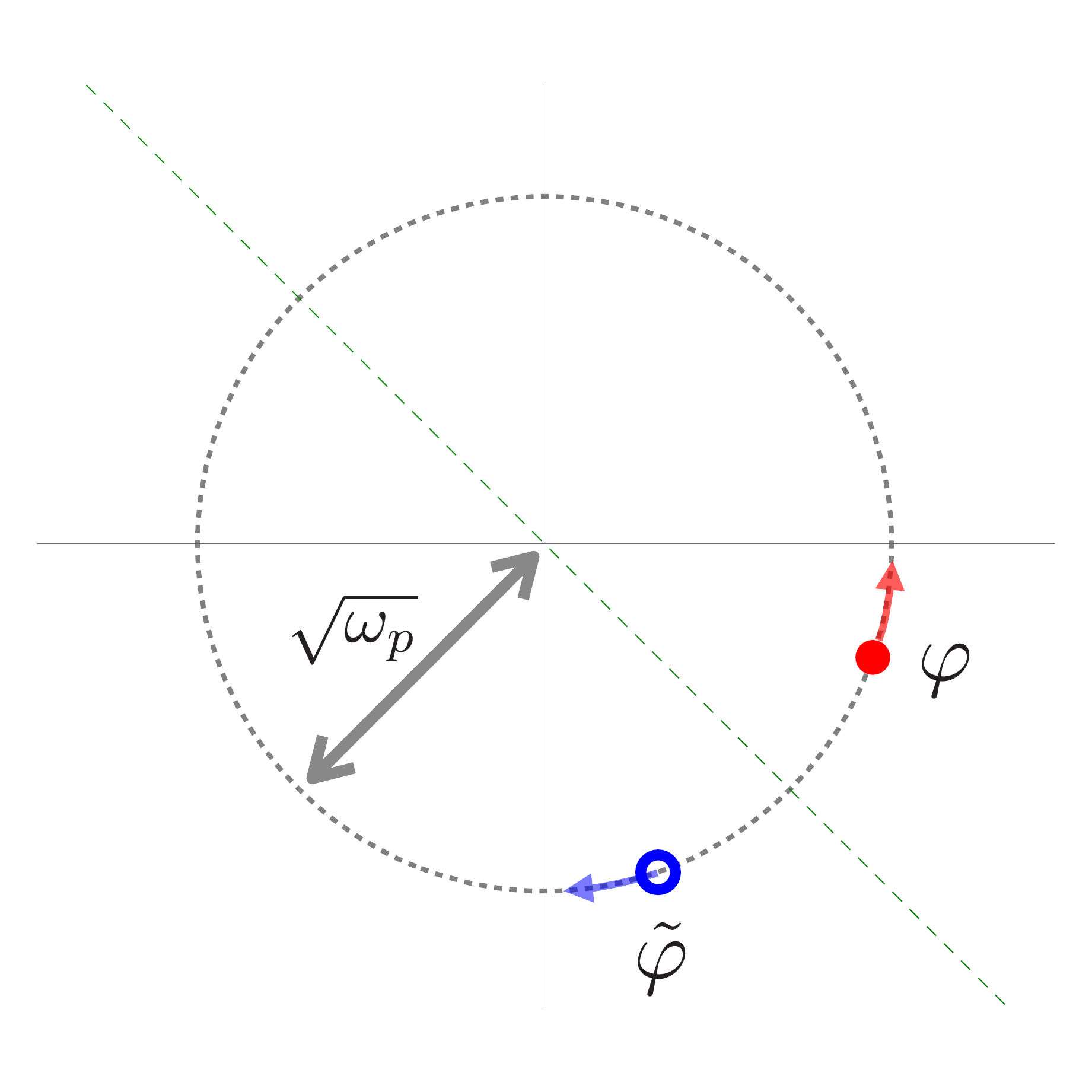}}
\label{fig:complexsolution}
\caption{Complex saddle point solution}
\end{figure}

In this section, we look for the saddle points 
responsible for the non-perturbative effects in this model. 
In the following, we use the rescaled variable 
$\varphi = \sqrt{g} \, \phi$ so that the action takes the form given in \eqref{eq:S0S1}. 
In addition to the classical vacuum solution $\varphi=0$, 
the classical action $S_E$ in Eq.\,\eqref{eq:S0S1} 
has non-trivial complex saddle point solutions. 
Such solutions can be found by complexifying the degree of freedom
\beq
S_E[\varphi, \bar \varphi] \rightarrow S_E[\varphi, \tilde \varphi], 
\eeq
where $S_E[\varphi, \tilde \varphi]$ is interpreted as 
a holomorphic functional of 
two independent complex variables 
$\varphi$ and $\tilde \varphi$. 
The saddle points can be found by solving 
the complexified equations of motion 
\beq
0 &=& \Big[+\p_\tau + i \mu + \tilde \varphi \varphi \Big] \varphi, \\
0 &=& \Big[ -\p_\tau + i \mu + \tilde \varphi \varphi \Big] \tilde \varphi.
\eeq
We can show by using the conservation laws 
that besides $\varphi = \tilde \varphi = 0$, 
there are infinitely many complex saddle points labeled 
by an integer $p \in \Z$
\beq
\varphi = \sqrt{-i \omega_p} \, \exp \left( \frac{2\pi i p \tau}{\beta} + i \theta \right), \hs{10}
\tilde \varphi = \sqrt{-i\omega_p} \, \exp \left( - \frac{2\pi i p \tau}{\beta} - i \theta \right),
\label{eq:sol_p}
\eeq
where $\theta$ is an integration constant (moduli parameter)
and $\omega_p = \mu + 2\pi p/\beta$ as in the previous section.
Note that $\tilde \varphi$ is not the complex conjugate of $\varphi$ (see Fig.\,\ref{fig:complexsolution}) and hence these solutions are complex saddle points 
which are not contained in the original configuration space before the complexification.  
The values of the action at these saddle points are given by
\beq
S_E = \frac{\beta \omega_p^2}{2g}. 
\eeq
We can confirm that these values agree with the non-perturbative exponents of the discontinuity of the perturbative part \eqref{eq:discon}. 
To compute the contributions from these complex saddle points,
let us consider the integration over the thimble $\mathcal J_p$
associated with the $p$-th complex saddle point
\beq
Z_p = \int_{\mathcal J_p} \D \varphi \, \exp \Big( -S_E[\varphi,\tilde \varphi] - S_W[\varphi,\tilde \varphi] \Big). 
\eeq
We first focus on the leading order contribution 
in the weak coupling limit $g \rightarrow 0$. 
Let $c_q$ and $\tilde c_q$ be the Fourier coefficients of $\varphi$ and $\tilde \varphi$ 
\beq
\varphi = \sqrt{g} \sum_{q=-\infty}^{\infty} c_q \exp \left( \frac{2\pi i q \tau}{\beta} \right), \hs{5}
\tilde{\varphi} = \sqrt{g} \sum_{q=-\infty}^{\infty} \tilde c_q \exp \left( \frac{2\pi i q \tau}{\beta} \right). 
\eeq
The $p$-th saddle point corresponds to 
the configuration with
\beq
c_q \ = \ e^{i \theta} \times \left\{ 
\begin{array}{cc} 
\sqrt{-i\omega_p/g} & \mbox{for $q=p$} \\
0 & \mbox{for $q \not = p$} 
\end{array} \right., 
\hs{10}
\tilde c_q \ = \ e^{-i \theta} \times \left\{ 
\begin{array}{cc} 
\sqrt{-i\omega_p/g} & \mbox{for $q=-p$} \\
0 & \mbox{for $q \not = -p$} 
\end{array} \right..
\eeq
Now let us consider the change of the integration variables from $(c_q,\tilde c_q)$ to a set of coordinates parameterizing the neighborhood of the saddle point in the configuration space. 
Choosing the new integration variables $b_q,\,\tilde b_q~(q=\pm1,\pm2,\cdots)$, $a$ and $\theta$ 
around the saddle point as\footnote{
The tangent directions corresponding to $b_q,\,\tilde b_q$ and $a$ are chosen so that their tangent vectors are orthogonal to the direction of the zero mode. The moduli integration over $0 < \theta < 2\pi$ compensate the missing zero mode direction.}
\beq
c_q \ = \ e^{i \theta} \times \left\{ 
\begin{array}{cc} 
\sqrt{-i\omega_p/g} + a & \mbox{for $q=p$} \\
b_{q-p} & \mbox{for $q \not = p$} 
\end{array} \right., 
\hs{10}
\tilde c_q \ = \ e^{-i \theta} \times \left\{ 
\begin{array}{cc} 
\sqrt{-i\omega_p/g} + a & \mbox{for $q=-p$} \\
\tilde b_{q+p} & \mbox{for $q \not = -p$} 
\end{array} \right.,
\eeq
we find that around the saddle point, 
the action $S_E$ and $S_W$ take the forms of
\beq
S_E = \beta \left[ \frac{\omega_p^2}{2 g} - 2 i \omega_p a^2 - i \sum_{q=1}^\infty \tilde B_{-q} X_{p,q} B_q \right] + \mathcal O(g^\frac{1}{2}), \hs{10}
S_W = p \pi i + \mathcal O(g^\frac{1}{2}), 
\eeq
with 
\beq
B_q = \ba{c} b_q \\ \tilde b_q \ea, \hs{5} 
\tilde B_{-q} = \ba{cc} \tilde b_{-q} \,, & b_{-q} \ea, \hs{5}
X_{p,q} = \ba{cc} \omega_{p-q} & \omega_p \\ \omega_p & \omega_{p+q} \ea.
\eeq
In the weak coupling limit, the integration measure takes the form of
\beq
N \prod_{q=-\infty}^\infty \frac{i}{2} dc_q \wedge d \tilde c_{-q} = N \sqrt{\frac{\omega_p}{ig}} \, da \wedge d\theta \, \prod_{q=1}^\infty \left( \frac{i}{2} db_q \wedge d \tilde b_{-q} \right) \wedge \left( \frac{i}{2} db_{-q} \wedge d \tilde b_{q} \right) + \mathcal O(g^0),
\eeq
where the normalization factor $N$ is the same 
as the one used to compute the perturbation series in Eq.\,\eqref{eq:normalization}. 
Using this integration measure, one can evaluate the leading order contribution from the $p$-th saddle point as 
\beq
Z_p \ = \ N \int \prod_{q=-\infty}^{\infty} \left(\frac{i}{2} d c_q \wedge d \tilde c_{-q} \right) \, \exp \left( - S_E - S_W \right) \ = \ N \, e^{-\frac{\beta \omega_p^2}{2 g} + p \pi i} \prod_{q=0}^\infty \mathcal I_{p,q} + \cdots,
\eeq
with
\beq
\mathcal I_{p,0} &=& \sqrt{\frac{\omega_p}{ig}} \int d a d\theta \, \exp \left( 2 i \omega_p \beta a^2 \right) \hs{20} = \ \pi \sqrt{\frac{2\pi}{\beta g}}, \\
\mathcal I_{p,q} &=& \int d b_q d \tilde b_q d b_{-q} d \tilde b_{-q} \, \exp \left( i \beta \tilde B_{-q} X_{p,q} B_q \right) 
\ = ~~~ \frac{1}{4q^2} \hs{15} (q \not = 0),
\eeq
where we have determined the integration contours
by the steepest descent method (Lefschetz thimble method). 
To evaluate the infinite product, 
let us consider the ratio between $Z_p$ and 
the leading order contribution around the perturbative vacuum $Z_0$
\beq
\frac{Z_p}{Z_0} = e^{-\frac{\beta \omega_p^2}{2 g} - \frac{i \beta \omega_p}{2}} \sqrt{\frac{2\pi}{\beta g}} i \beta \omega_p \prod_{q=1}^\infty \left[ 1 - \left( \frac{\beta \omega_p}{2\pi q} \right)^2 \right] + \cdots = e^{-\frac{\beta \omega_p^2}{2 g}} \sqrt{\frac{2\pi}{\beta g}} \left( 1 - e^{- i \beta \mu} \right) + \cdots,
\eeq
where we have used the fact that $Z_0$ in \eqref{eq:Z_0_infinite_product} can be rewritten as 
\beq
Z_0 = N e^{\frac{i \beta \mu}{2}} \prod_{q=0}^\infty \tilde{\mathcal I}_{p,q} , \hs{5} \mbox{with} \hs{5} 
\tilde{\mathcal I}_{p,q} = \left\{ \begin{array}{ll} \pi/(i \beta \omega_p) & \mbox{for $q=0$} \\ \pi^2/[(2\pi q)^2-(\beta \omega_p)^2] & \mbox{for $q \not = 0$} \end{array} \right..
\eeq
Since $Z_0 = (1-e^{-i\beta \mu})^{-1}$, 
we find that the leading order contribution of the $p$-th saddle point is given by
\beq
Z_p = \sqrt{\frac{2\pi}{\beta g}} \exp \left( - \frac{\beta}{2 g} \omega_p^2 \right) \Big[ 1 + \mathcal O(g) \Big]. 
\eeq
The higher order part can be determined 
by using the differential equation \eqref{eq:diff_eq}. 
Since the leading order part takes the form of the ``heat kernel", it satisfies the differential equation \eqref{eq:diff_eq}. 
As shown in Appendix \ref{appendix:diff_eq}, 
this is the unique solution that is regular at $\omega_p=0$. 
Therefore, there is no higher order correction,
i.e. $Z_p$ is one-loop exact 
\beq
Z_p = \sqrt{\frac{2\pi}{\beta g}} \exp \left(  - \frac{\beta}{2 g} \omega_p^2 \right). 
\label{eq:nonpert_contribution}
\eeq

\subsection{Intersection Numbers}\label{subsec:intersection}
Although we have determined the integral along the thimbles associated with the non-perturbative saddle points, 
not all of them contribute to the generating function $Z$. 
In the Lefschetz thimble method,   
the generating function can be constructed 
by combining the perturbation series and 
the non-perturbative contributions 
from the complex saddle points as
\beq
Z = Z_{\rm pert} + \sum_{p=-\infty}^\infty n_p Z_p,
\eeq
where $n_p$ is the intersection number between 
the original path integration contour and 
the dual thimble of the $p$-th saddle point. 
The dual thimble is defined as the set of points 
that flow to the $p$-th saddle point 
under the gradient flow of $S_E$   
\beq
\overline{\p_s \tilde \varphi} &=& \frac{1}{g} \, \Big[ + \p_\tau + i \mu + \tilde \varphi \varphi \Big] \varphi, \label{eq:flow1} \\
\overline{\p_s \varphi} &=& \frac{1}{g} \, \Big[ -\p_\tau + i \mu + \tilde \varphi \varphi \Big] \tilde \varphi, \label{eq:flow2}
\eeq
in the limit $s \rightarrow \infty$. 
The original integration contour is 
the subspace of the complexified configuration space 
specified by the condition $\tilde \varphi = \bar \varphi$ (complex conjugate of $\varphi$). 
If the original contour and the dual thimble intersect with each other, 
there exists a solution to the flow equation satisfying 
the following initial and final conditions 
with respect to the flow time $s \in [s_0, \infty)$ 
\beq
\tilde \varphi(s_0,\tau) = \bar \varphi(s_0,\tau), \hs{5}
\lim_{s \rightarrow \infty} (\varphi(s,\tau),\tilde \varphi(s,\tau)) = (\varphi_p(\tau),\tilde \varphi_p(\tau)),
\label{eq:ini_fin_cond}
\eeq
where $(\varphi_p(\tau),\tilde \varphi_p(\tau))$ denotes 
the $p$-th saddle points solution \eqref{eq:sol_p}. 
In the following, we determine the intersection number
by looking for a solution of the flow equation satisfying the condition \eqref{eq:ini_fin_cond}. 

\begin{figure}
\centering
\begin{tabular}{ccc}
\includegraphics[width=50mm, bb = 0 0 540 540]{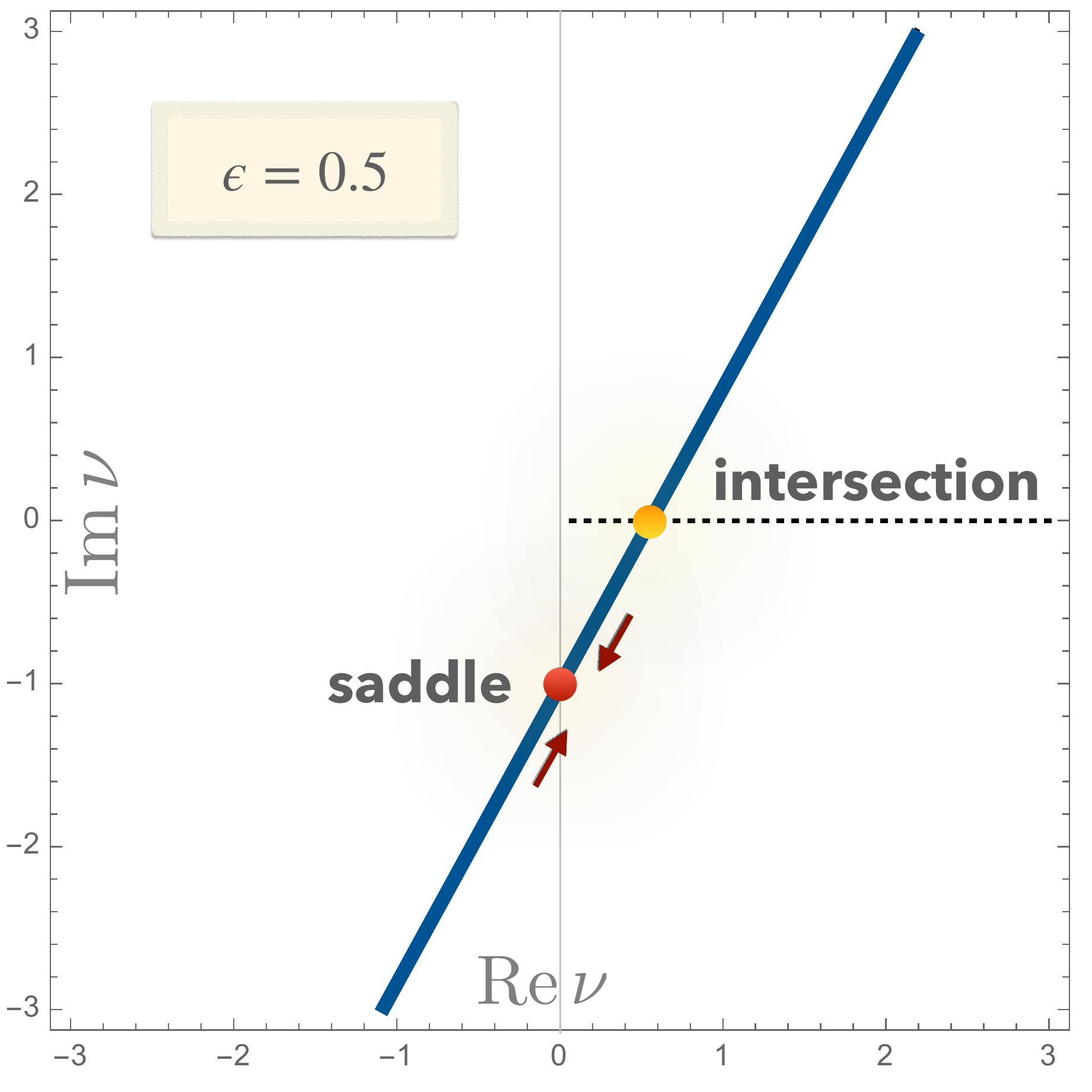}
&~\hs{20}~&
\includegraphics[width=50mm, bb = 0 0 540 540]{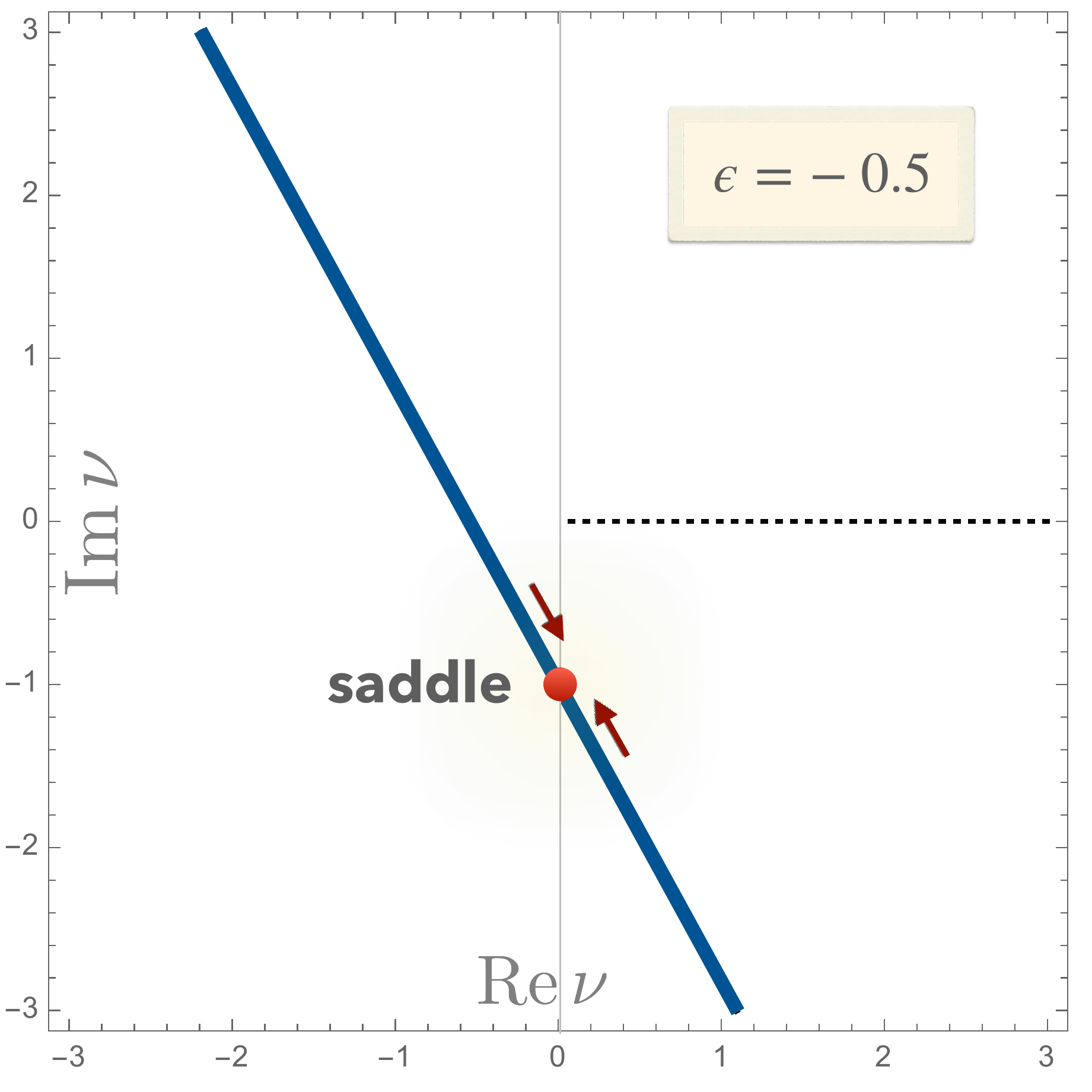}
\\ \\
(a) $\epsilon > 0$ & &
(b) $\epsilon < 0$
\end{tabular}
\caption{Flows in complex $\mathcal \nu$-plane ($\arg \omega_p >0$).}
\label{fig:intersection}
\end{figure}

Let us assume that the dual thimble intersects with the original integration contour at an isolated point\footnote{
This is a natural assumption since both the original integration contour and the dual thimble are half-dimensional subspaces of the complexified configuration space. 
If they have a higher dimensional intersection, 
we need to continuously deform the model 
so that the intersection becomes an isolated point.}. 
\begin{figure}
\centering
\begin{tabular}{ccc}
\includegraphics[width=60mm, bb = 0 0 530 530]{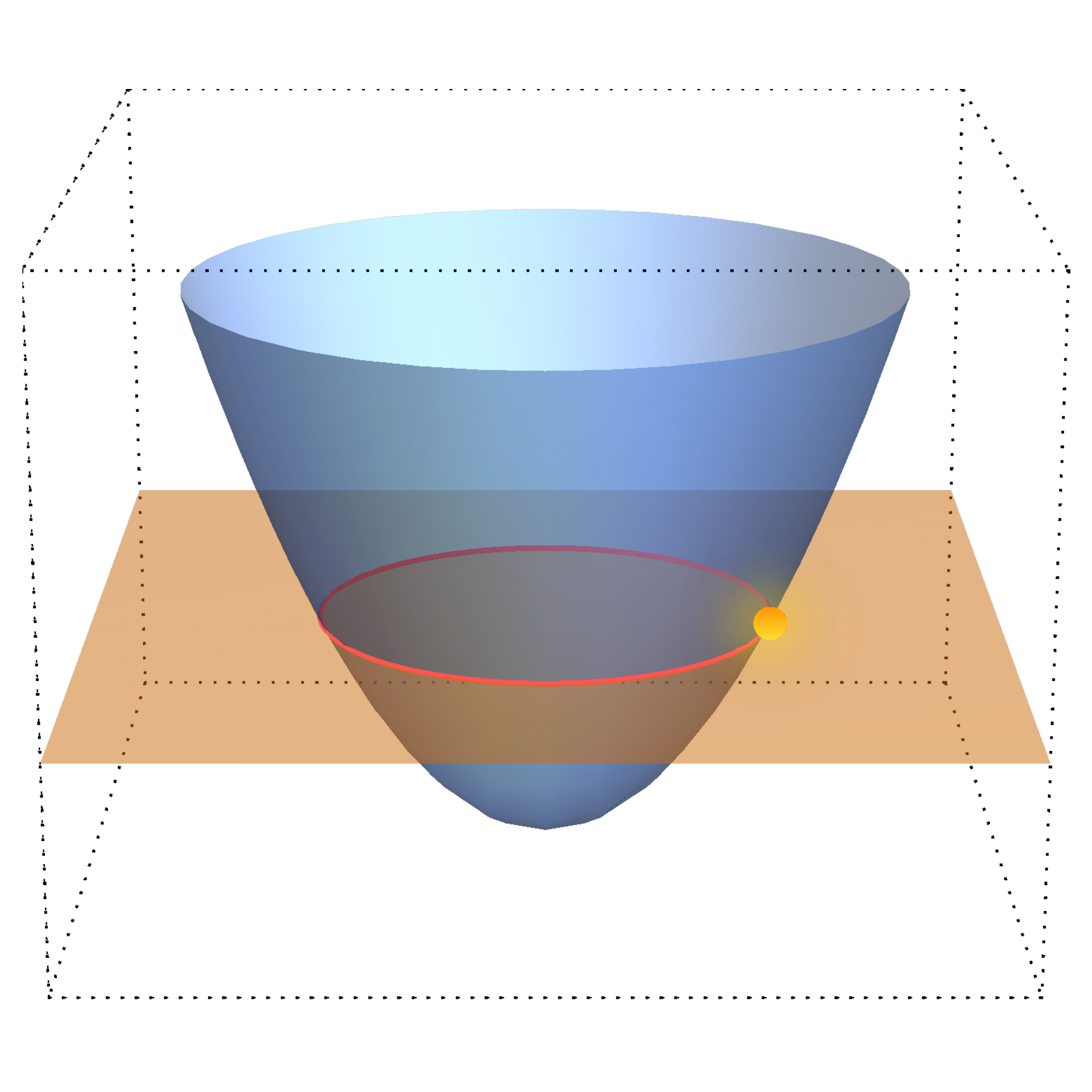} & {}\hs{20}{} &
\includegraphics[width=60mm, bb = 0 0 530 530]{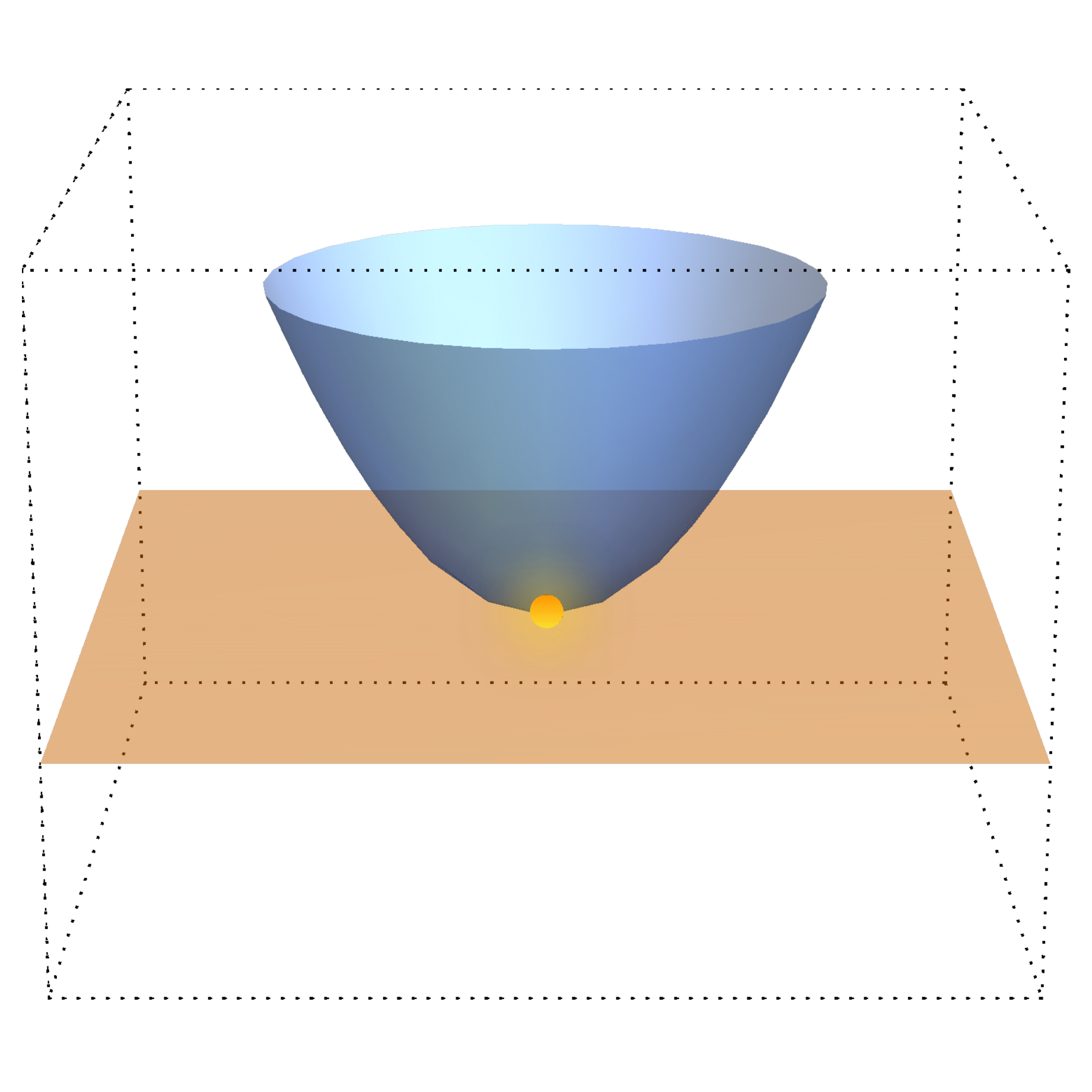} 
\end{tabular}
\caption{Intersection and symmetry. If two surfaces intersect at a point that is not a fixed point of a symmetry, the whole orbit is contained in the intersection of the surfaces (left panel). The intersection is a single point if it is a fixed point of the symmetry (right panel).}
\label{fig:my_label}
\end{figure}
Then, we can show that the intersection point must be 
at the fixed point of the simultaneous shift of 
the time and the angle variable 
\beq
\varphi(\tau) \rightarrow e^{\frac{2\pi i p \tau_0}{\beta}} \varphi(\tau-\tau_0), \hs{5}
\tilde \varphi(\tau) \rightarrow e^{-\frac{2\pi i p \tau_0}{\beta}} \tilde \varphi(\tau-\tau_0) \hs{5}
\mbox{with} ~~~ \tau_0 \in \R.
\eeq
The reason why the intersection is the fixed point is because the $p$-th saddle point is a fixed point of this symmetry and hence if the intersection point were not invariant under the symmetry, we have a continuous family of flow lines that intersects with the original contour along the orbit of the symmetry action (see Fig.\,\ref{fig:my_label}). This contradicts the assumption that the dual thimble intersects with the original integration contour at a single point.
Therefore, we assume the following invariant ansatz for the flow connecting the $p$-th saddle point and the intersection point 
\beq
\varphi = \sqrt{\nu(s)} \, e^{\frac{2\pi i p \tau}{\beta}+i\theta}, \hs{10}
\tilde \varphi = \sqrt{\tilde \nu(s)} \, e^{-\frac{2\pi i p \tau}{\beta} - i \theta},
\label{eq:flow_ansatz}
\eeq
where $\nu(s)$ and $\tilde \nu(s)$ are function depending only on the flow parameter $s$. 
By using the conservation law, 
we can show that $|\nu|-|\tilde \nu|$ is a constant on the flow 
\beq
\p_s \int d\tau \big( |\varphi|^2 - |\tilde \varphi|^2 \big) = \beta \p_s \big( |\nu|-|\tilde \nu| \big) = 0.
\eeq
Since $\nu(s)$ and $\tilde \nu(s)$ 
converge to the $p$-th saddle point value
for $s \rightarrow \infty$
\beq
\lim_{s \rightarrow \infty} \nu = \lim_{s \rightarrow \infty} \tilde \nu = -i \omega_p,
\eeq
the difference $ |\nu| - |\tilde \nu|$ vanishes 
for $s \rightarrow \infty$ and hence 
\beq
|\tilde \nu(s)| = |\nu(s)| \hs{5} \mbox{for all $s$}. 
\label{eq:nu_abs}
\eeq
Then, we find from the flow equations \eqref{eq:flow1} and \eqref{eq:flow2} that the difference of the arguments also vanishes
\beq
\p_s (\arg \nu - \arg \tilde \nu) = 0~~~\Longrightarrow~~~
\arg \tilde \nu(s) = \arg \nu(s) \hs{5} \mbox{for all $s$}.
\label{eq:nu_arg} 
\eeq
Using \eqref{eq:nu_abs} and \eqref{eq:nu_arg}, 
we can eliminate $\tilde \nu$ by setting $\tilde \nu = \nu$ and 
then the flow equations \eqref{eq:flow1} and \eqref{eq:flow2} reduce to a single equation for $\nu$
\beq
\nu'(s) = 2 |\nu(s)| \, \overline{\left[ \frac{\nu(s) + i \omega_p}{g} \right]}.
\eeq
This equation can be solved by using the conservation law
\beq
\p_s \, {\rm Im} \, \mathcal S_p(\nu) = 0 \hs{5} \mbox{with} \hs{5}
\mathcal S_p(\nu) \equiv \frac{\beta}{g} \left(\frac{1}{2} \nu^2+ i \omega_p \nu \right),
\label{eq:reduced_flow}
\eeq
where $\mathcal S_p(\nu)$ is the value of the action obtained 
by substituting the ansatz \eqref{eq:flow_ansatz} with 
$\tilde \nu = \nu$ 
into the original action $S_E$ in \eqref{eq:S0S1}. 
There are two solutions 
corresponding to the a pair of lines that 
flow to the saddle point from the opposite directions
\beq
\nu(s) = - i \omega_p \left(1 \pm \frac{e^{-i \epsilon} \coth m s}{\cosh m s \pm \cos \epsilon \, \sinh m s}\right),
\eeq
where we have defined
\beq
\epsilon = \arg \omega_p - \frac{1}{2} \arg g, \hs{10}
m = \left| \frac{2\omega_p}{g} \right|.
\eeq
To examine if this flow intersects the original integration contour,  
it is convenient to see the orbit of the Noether charge $\nu = \varphi \tilde \varphi$ in the complex plane. 
We can show that the following relation holds along the flow
\beq
{\rm Re} \left( e^{i \epsilon} \nu / \omega_p \right) \ = \ \sin \epsilon.
\label{eq:N_line}
\eeq
Therefore, this flow is a straight line on the complex $\nu$-plane (see Fig.\,\ref{fig:intersection}). 
Since ${\rm Im} \, \nu = 0 $ and ${\rm Re} \, \nu > 0$ on the original integration contour ($\tilde \varphi = \bar \varphi$), 
an intersection point exists only when the line \eqref{eq:N_line} intersects the positive real axis on the complex $\nu$-plane, 
that is, if the parameters satisfy the condition
\beq
{\rm Re} \, \nu \, |_{\, {\rm Im} \, \nu = 0} \ = \ \frac{|\omega_p|\sin \epsilon}{\cos (\arg g/2)} \ > \ 0,
\eeq
there is an intersection 
between the original contour and the dual thimble.
Therefore, the intersection number is given by
\beq
n_p \ = \ \left\{ \begin{array}{cc} 1 & \mbox{for $\sin \epsilon > 0$} \\
0 & \mbox{for $\sin \epsilon < 0$} \end{array} \right. \, = \ \frac{1}{2} \left[ 1 + {\rm sign} \left( {\rm Im} \sqrt{\frac{\beta}{2g}} \omega_p \right) \right],
\label{eq:intersection_n}
\eeq
where we have assumed that 
$\arg g$ is small and hence 
$\cos(\arg g/2) > 0$. 

\subsection{Exact Generating Function and Comparison with Operator Formalism}\label{subsec:canonical}
Having determined 
the perturbation series \eqref{eq:P.S.},
all the non-perturbative contributions \eqref{eq:nonpert_contribution} and
the intersection numbers \eqref{eq:intersection_n}, 
we can construct the transseries for the generating function 
by combining them as
\beq
Z = Z_{\rm pert} + \sum_{p=-\infty}^\infty n_p Z_p = \sum_{n=0}^{\infty} \frac{1}{n!} \left( \frac{g}{2\beta} \frac{\p^2}{\p \mu^2} \right)^n \frac{1}{1-e^{-i \beta \mu}} + \sqrt{\frac{2\pi}{\beta g}}\sum_{p=-\infty}^{\infty} n_p \, \exp \left( -\frac{\beta}{2 g} \omega_p^2 \right). 
\eeq
By applying the Borel resummation 
to the perturbation series as in Eq.\,\eqref{eq:Z_pert}
and taking into account the discontinuities of the intersection numbers \eqref{eq:intersection_n},  
we can write down the unambiguous form of 
the full generating function as
\beq
Z  =  \frac{1}{2} - \sqrt{\frac{\pi}{2\beta g}} \sum_{p=-\infty}^{\infty} e^{- \frac{\beta}{2g} \omega_p^2} \left[ {\rm erf} \left( i \sqrt{\frac{\beta}{2g}} \omega_p \right) - 1 \right].
\label{eq:G.F._thimble}
\eeq
This shows that the ambiguities of the perturbative and non-perturbative sectors completely cancel out each other 
in the transseries obtained through the Lefschetz thimble method. 

We can show that the expression \eqref {eq:G.F._thimble} 
is not only well-defined 
but also exact by comparing it with 
the generating function obtained in the operator formalism.
By using the number eigenstates, 
the generating function can be written as 
\beq
Z \ = \ \tr \left[ e^{-\beta(\hat H + i \mu \hat{\mathcal N)}} \right] \ = \ \sum_{n=0}^\infty \exp \left[ - \frac{\beta g}{2} n^2 - i \beta \mu n \right]. 
\eeq
To compare this with the result of 
the Lefschetz thimble method
\eqref{eq:G.F._thimble}, 
let us use the relation
\beq
\sum_{n=0}^\infty f(g n) ~=~ \frac{1}{2} f(0) + \frac{1}{g} \sum_{p=-\infty}^{\infty} \int_0^\infty d\nu \, f(\nu) \, e^{-\frac{2\pi i p \nu}{g}}.
\label{eq:poisson}
\eeq
This relation can be regarded 
as a variant of the Poisson resummation. 
Applying this resummation method, 
we can rewrite the generating function as
\beq
Z = \frac{1}{2} + \frac{1}{g} \sum_{p=-\infty}^{\infty} \int_0^\infty d\nu \, \exp \left[ - \mathcal S_p(\nu) \right] \hs{5} 
\mbox{with} ~~~ \mathcal S_p = \frac{\beta}{g} \left( \frac{\nu^2}{2} + i \omega_p \nu \right).
\label{eq:G.F._operator}
\eeq
Evaluating the integrals by using the definition of the error function \eqref{eq:error}, 
we find the complete agreement of the generating functions
obtained through the Lefschetz thimble method \eqref{eq:G.F._thimble} and the operator formalism \eqref{eq:G.F._operator}. 

\begin{figure}
\centering
\begin{tabular}{ccc}
\includegraphics[width=50mm, bb = 0 0 540 540]{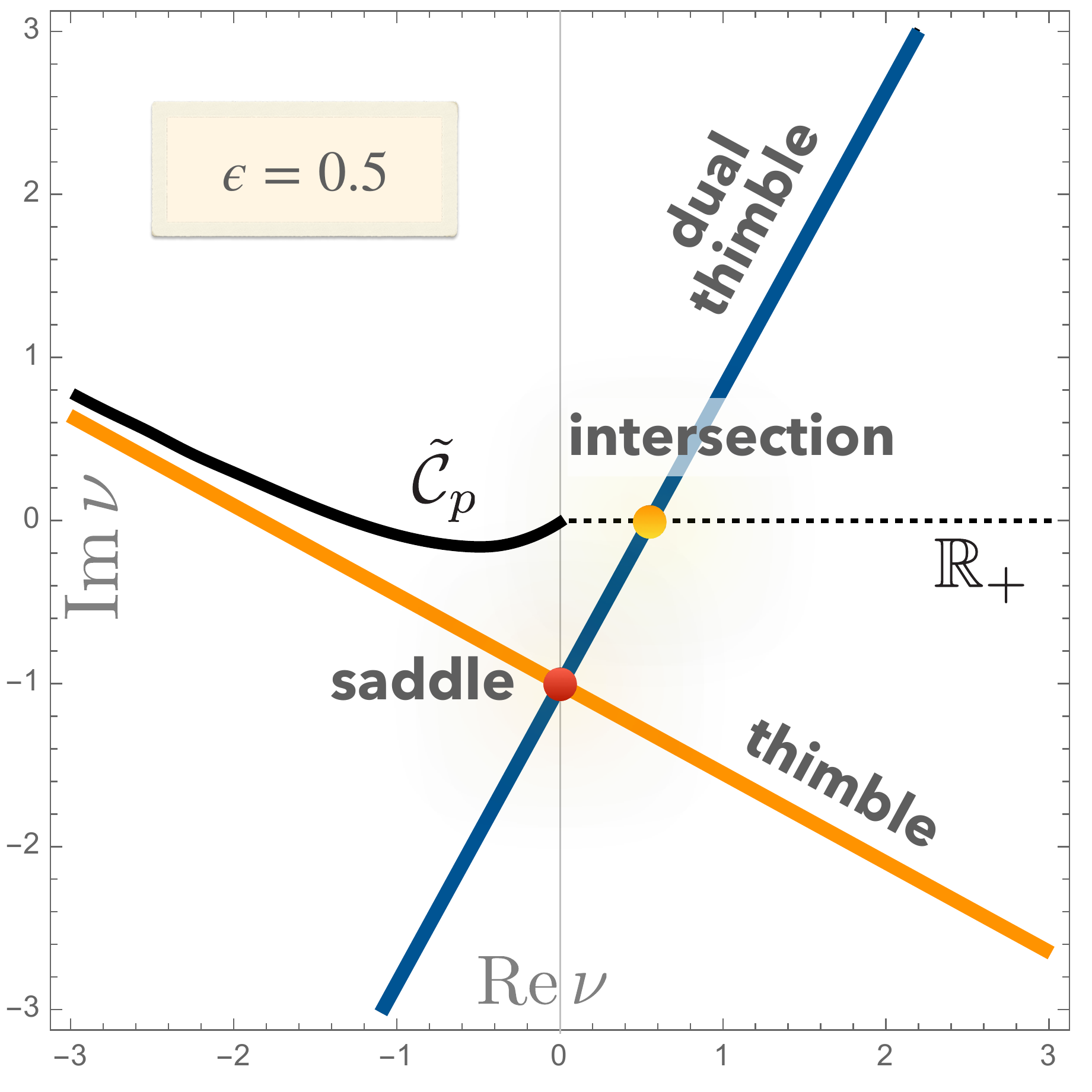}
&~\hs{20}~&
\includegraphics[width=50mm, bb = 0 0 540 540]{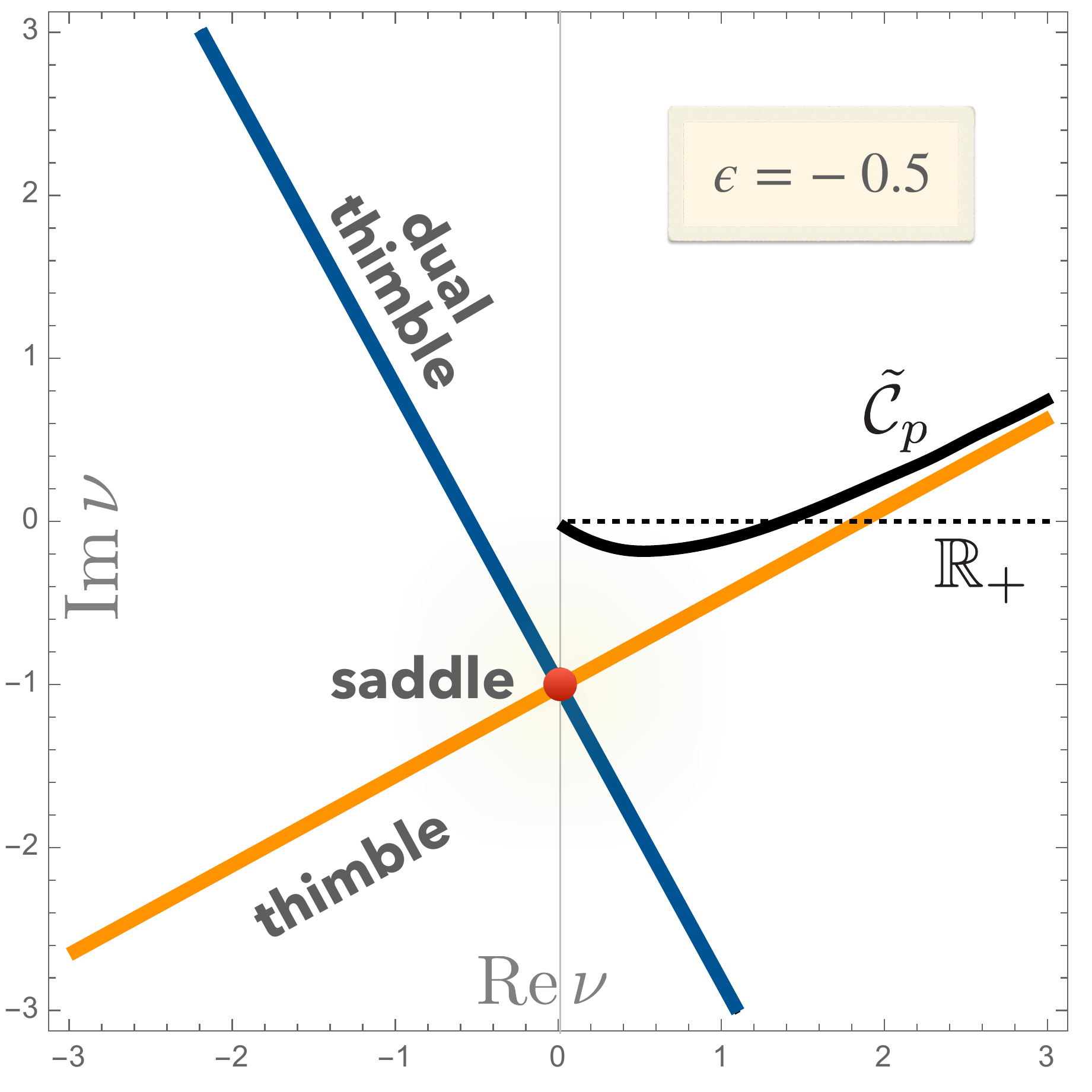}
\\ \\
(a) $\epsilon > 0$ & &
(b) $\epsilon < 0$
\end{tabular}
\caption{Thimbles in complex $\mathcal \nu$-plane ($\arg \omega_p >0$). (a) The positive real axis $\R_{\geq 0} $ can be decomposed into the steepest ascent path $\tilde{\mathcal C}_p$ and the thimble associated with the saddle point at $\nu=-i\omega_p$. (b) The positive real axis $\R_{\geq 0} $ can be continuously deformed to $\tilde{\mathcal C}_p$.}
\label{fig:thimbles}
\end{figure}

It is worth examining how the transseries 
for the generating function is obtained 
from the viewpoint of the operator formalism. 
To extract the perturbation series from \eqref{eq:G.F._operator}, 
let us consider steepest ascent path $\tilde{\mathcal C}_p$ 
of $\mathcal S_p(\nu)$ starting from the origin in the complex $\nu$-plane and decompose the integral along the positive real axis $\R_{\geq 0} $ as 
\beq
\frac{1}{g} \int_{\mathbb R_+} d\nu \, \exp [-\mathcal S_p(\nu)] = \frac{1}{g} \int_{\tilde{\mathcal C}_p} d\nu \, \exp [-\mathcal S_p(\nu)] + \frac{1}{g} \int_{\mathbb R_+-\tilde{\mathcal C}_p} d\nu \, \exp [-\mathcal S_p(\nu)], 
\label{eq:nu_int_decomposed}
\eeq
where $\mathbb R_+-\tilde{\mathcal C}_p$ is the path consisting of the positive real axis and the inverse path of $\tilde{\mathcal C}_p$ connected at the origin (see Fig.\,\ref{fig:thimbles}). 
We can show that the first term gives the perturbative part 
by changing the variable as 
\beq
\frac{1}{g} \int_{\tilde{\mathcal C}_p} d\nu \, \exp [-\mathcal S_p(\nu)] = - \frac{i}{\beta \omega_p} \int_0^t dt \, \frac{e^{-t}}{\sqrt{1-\frac{2gt}{\beta \omega_o^2}}} \hs{5} \mbox{with} \hs{5} t = \mathcal S_p(\nu).
\eeq
Summing over $p$, 
we find that the collection of these terms 
and $1/2$ in \eqref{eq:G.F._operator} 
correspond to the Borel resummation of the perturbation series \eqref{eq:pert_Borel}. 
The second integral in \eqref{eq:nu_int_decomposed}
can be evaluated by applying the Lefschetz thimble method to this integral. 
The saddle point of $\mathcal S_p(\nu)$ is located at 
$\nu = - i \omega_p$, 
which is nothing but the value of $\nu = \varphi \tilde \varphi$
for the $p$-th saddle point of the original action $S_E$ in Eq.\,\eqref{eq:sol_p}. 
Evaluating the integral along the associated thimble, 
we find that the saddle contribution agrees with \eqref{eq:nonpert_contribution}. 
The dual thimble is the path determined from 
${\rm Im} \, \mathcal S_p(\nu) = {\rm Im} \, \mathcal S_p(-i\omega)$. 
This agrees with the flow determined by the equation \eqref{eq:reduced_flow} reduced from the original flow equations.
Therefore, we obtain the same intersection numbers as Eq.\,\eqref{eq:intersection_n}. 
In this way, we can see the agreement of the transseries 
obtained from the path integral and operator formalism 
through the thimble analysis of 
the single variable functions $\mathcal S_p(\nu)$.

\section{Generalization to Arbitrary $U(1)$ Symmetric Potential}\label{sec:generalization}

\subsection{Preliminary}
In the previous section, 
we have seen that the Lefschetz thimble method 
gives exact results in the case of the quartic potential. 
It is also possible to generalize the discussion to the case of an arbitrary $U(1)$ symmetric potential
\beq
L ~=~ i \bar \phi \p_t \phi + \frac{1}{g} V(g |\phi|^2) ~=~ \frac{1}{g} \Big[ i \bar \varphi \p_t \varphi + V(|\varphi|^2) \Big], \hs{10} (\varphi = \sqrt{g} \, \phi).
\eeq
In the following, we assume that 
the potential $V(|\varphi|^2)$ has 
its minimum at $\varphi = 0$
and can be expanded as
\beq
V( |\varphi|^2) \ = \ \sum_{n} \kappa_n  |\varphi|^{2n} \ = \ \kappa_1 |\varphi|^2 + \kappa_2 |\varphi|^4 + \cdots. 
\eeq
As in the previous case, we consider the generating function
\beq
Z = \tr \left[ e^{-\beta (\hat H + i \mu \hat{\mathcal N})} \right],
\label{eq:Z_def_general}
\eeq
where the Hamiltonian $\hat H$ and the conserved charge $\hat{\mathcal N}$ are given by
\beq
\hat H = \frac{1}{g} V(g \hat{\mathcal N}), \hs{10} \hat{\mathcal N} = \hat \phi^\dagger \hat \phi. 
\eeq
We can show that the generating function satisfies the differential equation
\beq
\left[ - \frac{1}{\beta} \frac{\p}{\p g} + \frac{1}{g^2} V \left( g \frac{i}{\beta} \frac{\p}{\p \mu} \right) - \frac{1}{g} \frac{i}{\beta} \frac{\p}{\p \mu} V' \left( g \frac{i}{\beta} \frac{\p}{\p \mu} \right) \right] Z \ = \ 0. 
\label{eq:diff_eq_general}
\eeq
We will use this differential equation 
to determine the perturbation series in the following.

From the viewpoint of operator formalism, 
the generating function can be calculated 
by using the number eigenstates $\hat{\mathcal N} | n \rangle = n |n \rangle$ as 
\beq
Z = \sum_{n=0}^\infty \exp \left[ - \frac{\beta}{g} V(g n) - i \beta \mu n \right].
\eeq
On the other hand, 
the path integral expression for
the generating function is given by
\beq
Z = \int \mathcal D \phi \, \exp \left( - S_E - S_W \right),
\eeq
where $S_E$ is the classical Wick-rotated action
\beq
S_E = \frac{1}{g} \int_0^\beta d\tau \Big[ \bar \varphi \p_\tau \varphi +  V(|\varphi|^2) + i \mu |\varphi|^2 \Big],
\eeq
and $S_W$ is the part generated when the original Hamiltonian is rewritten in terms of the Weyl ordered operators
\beq
S_W = \int_0^\beta d\tau \left[ - \frac{1}{2} V'(|\varphi|^2) - \frac{i\mu}{2} + \mathcal O(g) \right].
\eeq
To derive this expression, we have used 
\beq
\hat{\mathcal N}^n = (\hat{\mathcal N})^n_W - \frac{n}{2} (\hat{\mathcal N})^{n-1}_W +  \mathcal O(g),
\eeq
where $(\hat{\mathcal N})^n_W$ stands 
for the Weyl ordered operator defined in \eqref{eq:Weyl}.
We will not use the details of the higher order terms
since their contributions can be determined 
through the differential equation.  

\subsection{Perturbation Series}
Let us first consider perturbative expansion of the generating function $Z$ with respect to the coupling constant $g$. 
Substituting the power series ansatz 
\beq
Z_{\rm pert} = \sum_{k=0}^\infty C_k \, g^k,
\eeq
into the differential equation \eqref{eq:diff_eq_general}, 
we obtain to a recursion relation
from which the coefficients $C_k$ can be determined 
order-by-order as 
\beq
C_{k} = - \beta \sum_{l=1}^{k} \frac{l}{k} \kappa_{l+1} \left( \frac{i}{\beta} \frac{\p}{\p \mu} \right)^{l+1} C_{k-l},
\eeq
where the initial term is given 
by the generating function in the free theory 
\beq
C_0 = \frac{1}{1-e^{- \beta (\kappa_1 + i \mu)}}.
\label{eq:init_cond}
\eeq
We can show that the Borel resummation of the perturbation series is given by
\beq
Z_{\rm pert} = \int_0^\infty dt \, e^{-t} \left( \frac{1}{2} + \frac{1}{\beta} \sum_{p=-\infty}^\infty \frac{1}{V'(\nu_p(gt))+i \omega_p} \right),
\label{eq:Borel_general}
\eeq
where $\omega_p = \mu + \frac{2\pi p}{\beta}$ and 
$\nu_p(g t)$ is the solution of the equation 
\beq
\mathcal S_p(\nu) \ \equiv \ \frac{\beta}{g} \left[ V(\nu) + i \omega_p \nu \right] \ = \ t \hs{5} \mbox{with} ~~~~ \nu(0)=0. 
\label{eq:changeofvariable_nu_t}
\eeq 
We can check that 
Eq.\,\eqref{eq:Borel_general} gives 
the correct perturbation series by confirming 
that it satisfies the differential equation \eqref{eq:diff_eq_general}. 
For this purpose it is convenient to change 
the integration variable from $t$ to $\nu$ as
\beq
Z_{\rm pert} = \frac{1}{2} + \sum_{p=-\infty}^\infty \int_{\widetilde{\mathcal C}_p} \frac{d\nu}{g} \, \exp \left( - \mathcal S_p(\nu) \right),
\label{eq:pert_nu}
\eeq
where the integration contour 
$\widetilde{\mathcal C}_p$ is the image of the positive real axis under the map from the $t$-plane to the $\nu$-plane, 
that is, the ascending flow of $\mathcal S_p(\nu)$
emanating from the origin on the complex $\nu$-plane (see the examples in Fig.\,\ref{fig:general_thimble}). 
Substituting into \eqref{eq:diff_eq_general}, 
we find that \eqref{eq:pert_nu} satisfies the differential equation
\beq
\left[ - \frac{1}{\beta} \frac{\p}{\p g} + \frac{1}{g^2} V \left( g \frac{i}{\beta} \frac{\p}{\p \mu} \right) - \frac{1}{g} \frac{i}{\beta} \frac{\p}{\p \mu} V' \left( g \frac{i}{\beta} \frac{\p}{\p \mu} \right) \right] Z_{\rm pert} = \sum_{p=-\infty}^\infty \int_{\widetilde{\mathcal C}_p} \frac{d\nu}{g} \, \p_\nu \left[ \frac{\nu}{g \beta} e^{-\mathcal S_p} \right] = 0. 
\label{eq:diff_eq_check}
\eeq
Furthermore, 
\eqref{eq:pert_nu} satisfies the initial condition 
\eqref{eq:init_cond} and 
hence $Z_{\rm pert}$ in \eqref{eq:Borel_general} 
gives the correct perturbation series.

The formal Borel resummation \eqref{eq:Borel_general} 
of the perturbation series is non-Borel summable 
if the integrand (Borel transform) has singularities along the positive real axis in the Borel plane (complex $t$-plane). 
This occurs when 
one of the contours $\widetilde{\mathcal C}_p$ in Eq.\,\eqref{eq:pert_nu}
connects the origin and a saddle point of $\mathcal S_p(\nu)$, 
that is, a point at which $\nu$ satisfies
\beq
\mathcal S_p'(\nu) \propto V'(\nu) + i \omega_p = 0. 
\label{eq:cond_sing}
\eeq  
Although such singularities can be avoided 
by complexifying the coupling constant 
$g \rightarrow |g| e^{\pm i 0}$, 
the Borel resummation of the perturbation series 
has ambiguities of the form
\beq
\Delta Z_{\rm pert} = Z_{\rm pert}^{(+)} - Z_{\rm pert}^{(-)} = \sum_{p=-\infty}^\infty \left[ \int_{\widetilde{\mathcal C}_p^{(+)}} \frac{d\nu}{g} \, \exp \left( - \mathcal S_p(\nu) \right) - \int_{\widetilde{\mathcal C}_p^{(-)}} \frac{d\nu}{g} \, \exp \left( - \mathcal S_p(\nu) \right) \right], 
\eeq 
where $\widetilde{\mathcal C}_p^{(\pm)}$ 
are the ascending flows of $\mathcal S_p(\nu)$ 
for ${\rm arg} \, g = \pm 0$. 
Noting that 
$\widetilde{\mathcal C}_p^{(+)}-\widetilde{\mathcal C}_p^{(-)}$ 
is the thimble associated with the saddle point 
connected to the origin by the flow 
$\tilde{\mathcal C}_p |_{\arg g =0}$ (see Fig.\,\ref{fig:general_thimble}-(b)), 
we can rewrite the discontinuity as 
\beq
\Delta Z_{\rm pert} \ = \ \sum_{p,\sigma} m_{p,\sigma} (\Delta Z_{\rm pert})_{p,\sigma} \ = \ \sum_{p,\sigma} m_{p,\sigma} \int_{\mathcal J_{p,\sigma}} \frac{d\nu}{g} \, \exp \left( - \mathcal S_p(\nu) \right),
\label{eq:ambiguity_general}
\eeq
where 
$\sigma$ is the label of the saddle points of 
$\mathcal S_p(\nu)$, 
$\mathcal J_{p,\sigma}$ is the thimble\footnote{
The orientation of the thimble is chosen so that 
\beq
\int_{\mathcal J_{p,\sigma}} \frac{d\nu}{g} \, \exp \left( - \mathcal S_p(\nu) \right) = \sqrt{\frac{2\pi}{\beta g \mathcal S_p''(\nu_{p,\sigma})}} \exp \left( - \mathcal S_p(\nu_{p,\sigma}) \right) \Big[ 1 + \mathcal O(g) \Big]. \notag
\eeq}
associated with the saddle point $\nu_{p,\sigma}$
and the coefficient $m_{p,\sigma}$ is given by
\beq
m_{p,\sigma} \ = \ 
\left\{ 
\begin{array}{cc}
{\rm sign} ( {\rm Im} \, \mathcal S_p'(0) ) & ~~~ \mbox{if $\nu_{p,\sigma} \in \tilde{\mathcal C}_p$} \\
0 & ~~~ \mbox{if $\nu_{p,\sigma} \not \in \tilde{\mathcal C}_p$}
\end{array} \right.. 
\eeq
Note that each $(\Delta Z_{\rm pert})_{p,\sigma}$ satisfies 
the differential equation \eqref{eq:diff_eq_general}. 
In the next section, we will see that these ambiguities are canceled by the contributions from complex saddle point solutions. 


\subsection{Complex Saddle Points}
Let us look for the saddle points that cancel 
the ambiguities of the perturbation series in  Eq.\,\eqref{eq:ambiguity_general}. 
The complexified equations of motion $\delta S_E/\delta \varphi = \delta S_E/\delta \tilde \varphi = 0$ are given by
\beq
0 &=& \Big[ + \p_\tau + i \mu + V'(\tilde \varphi \varphi) \Big] \varphi, \\
0 &=& \Big[ - \p_\tau + i \mu + V'(\tilde \varphi \varphi) \Big] \tilde \varphi. 
\eeq
Using the conservation law, 
we can show that the solution takes the form
\beq
\varphi = \sqrt{\nu} \exp \left( \frac{2\pi p i \tau}{\beta} + i \theta \right), \hs{10}
\tilde \varphi = \sqrt{\nu} \exp \left( - \frac{2\pi p i \tau}{\beta} - i \theta \right),
\eeq
where $\nu$ is a constant satisfying the condition 
\beq
V'(\nu) + i \omega_p = 0. 
\label{eq:cond_sing2}
\eeq
This is nothing but the condition in Eq.\,\eqref{eq:cond_sing} 
that determines the locations of the singularities 
in the Borel plane for the perturbation series in 
Eq.\,\eqref{eq:Borel_general}.  
Suppose that $\nu_{p,\sigma}$ is a solution of \eqref{eq:cond_sing2}.
Then, we can show that the value of action for the solution corresponding to $\nu_{p,\sigma}$ is given by
\beq
\mathcal S_p(\nu_{p,\sigma}) = \frac{\beta}{g} \left[ V(\nu_{p,\sigma}) + i \omega_p \nu_{p,\sigma} \right]. 
\eeq

The leading order contributions from these saddle points can be calculated similarly to the previous case. 
For example, we can show that the one-loop determinant can be obtained 
by replacing $g$ in the previous section to $g V''(\nu_{p,\sigma})$
\beq
(-1)^p \sqrt{\frac{2\pi}{\beta g}} ~~ \longrightarrow ~~ (-1)^p \sqrt{\frac{2\pi}{\beta g V''(\nu_{p,\sigma})}}.
\eeq
The leading order part of $S_W$ is given by
\beq
S_W = - \frac{1}{2} \left[ \beta V'(\nu) + i \beta \mu \right] + \mathcal O(g) = \pi p i + \mathcal O(g),
\eeq
and hence the leading order part of the saddle point contribution takes the form of 
\beq
Z_{p,\sigma} = \sqrt{\frac{2\pi}{\beta g V''(\nu_{p,\sigma})}} \, e^{-\mathcal S_p(\nu_{p\sigma})} \Big[ 1 + \mathcal O(g) \Big]. 
\eeq
This leading order contribution is identical 
to that of the corresponding ambiguity of 
the perturbation series in Eq.\,\eqref{eq:ambiguity_general}.
Since the higher order part can be uniquely determined 
from the leading part by the differential equation, 
the agreement of the leading order parts implies that 
the saddle point contribution $Z_{p,\sigma}$ and 
the corresponding ambiguity 
$(\Delta Z_{\rm pert})_{p,\sigma}$ in Eq.\,\eqref{eq:ambiguity_general} agree
to all orders in the coupling constant $g$
\beq
Z_{p,\sigma} = (\Delta Z_{\rm pert})_{p,\sigma} = \int_{\mathcal J_{p,\sigma}} \frac{d\nu}{g} \, \exp \left( - \mathcal S_p(\nu) \right). 
\label{eq:non_pert_all}
\eeq
Therefore, the transseries 
\beq
Z = \int_0^\infty dt \, e^{-t} \left( \frac{1}{2} + \frac{1}{\beta} \sum_{p=-\infty}^\infty \frac{1}{V'(\nu_p(gt))+i \omega_p} \right) + \sum_{p,\sigma} n_{p,\sigma} Z_{p,\sigma},
\label{eq:transseries_general}
\eeq
do not have ambiguities if the intersection numbers 
$n_{p,\sigma}$ have appropriate discontinuities 
$\Delta n_{p,\sigma} = \pm 1$ at $\arg g = 0$. 
In the next section, 
we determine the intersection numbers 
$n_{p,\sigma}$ by using the flow equation. 

\subsection{Intersection numbers}\label{subsec:intersection_general}
\begin{figure}[!ht]
\centering
\begin{tabular}{ccc}
\fbox{
\includegraphics[width=50mm, bb = 15 0 535 535]{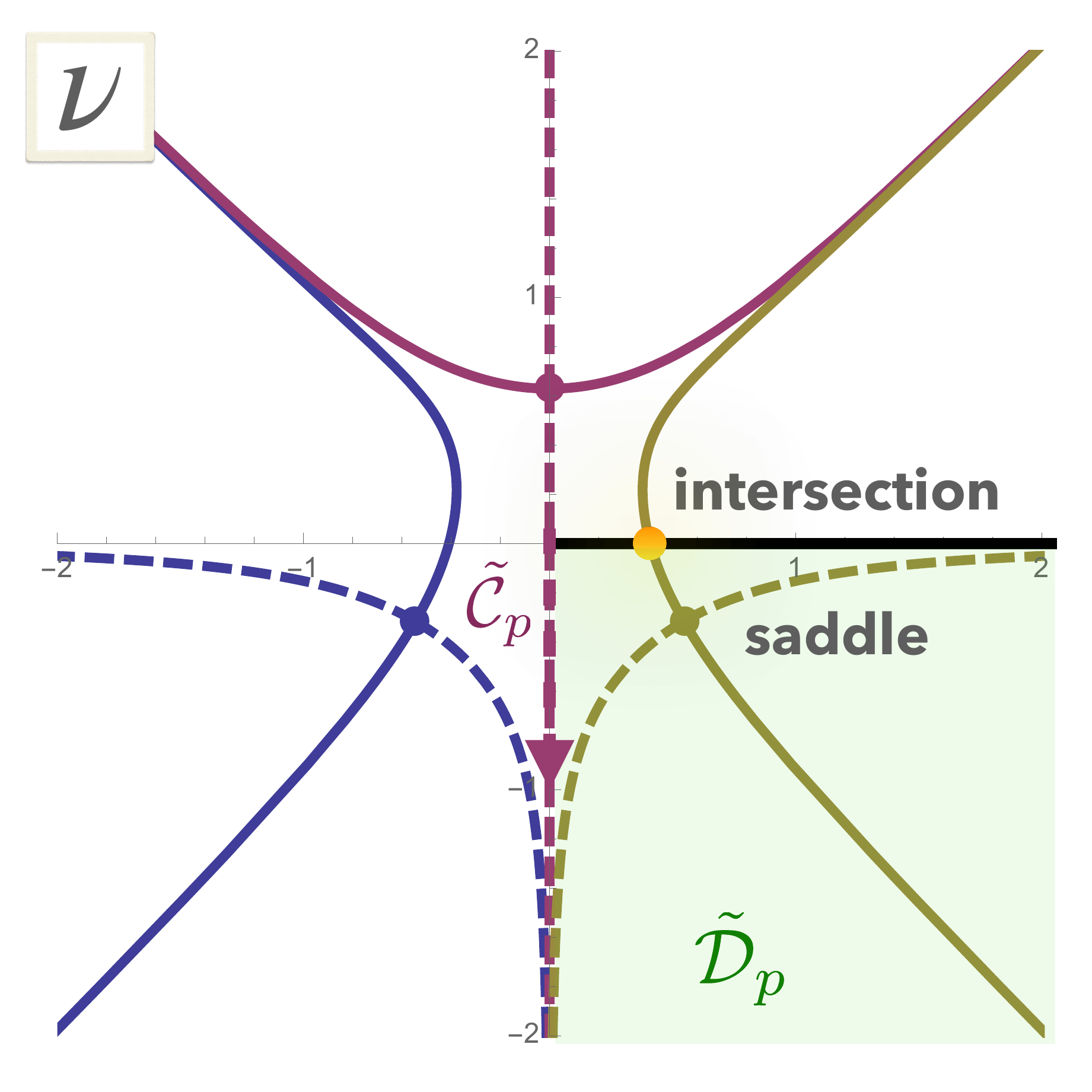}}
&~\hs{20}~&
\fbox{
\includegraphics[width=50mm, bb = 15 0 535 535]{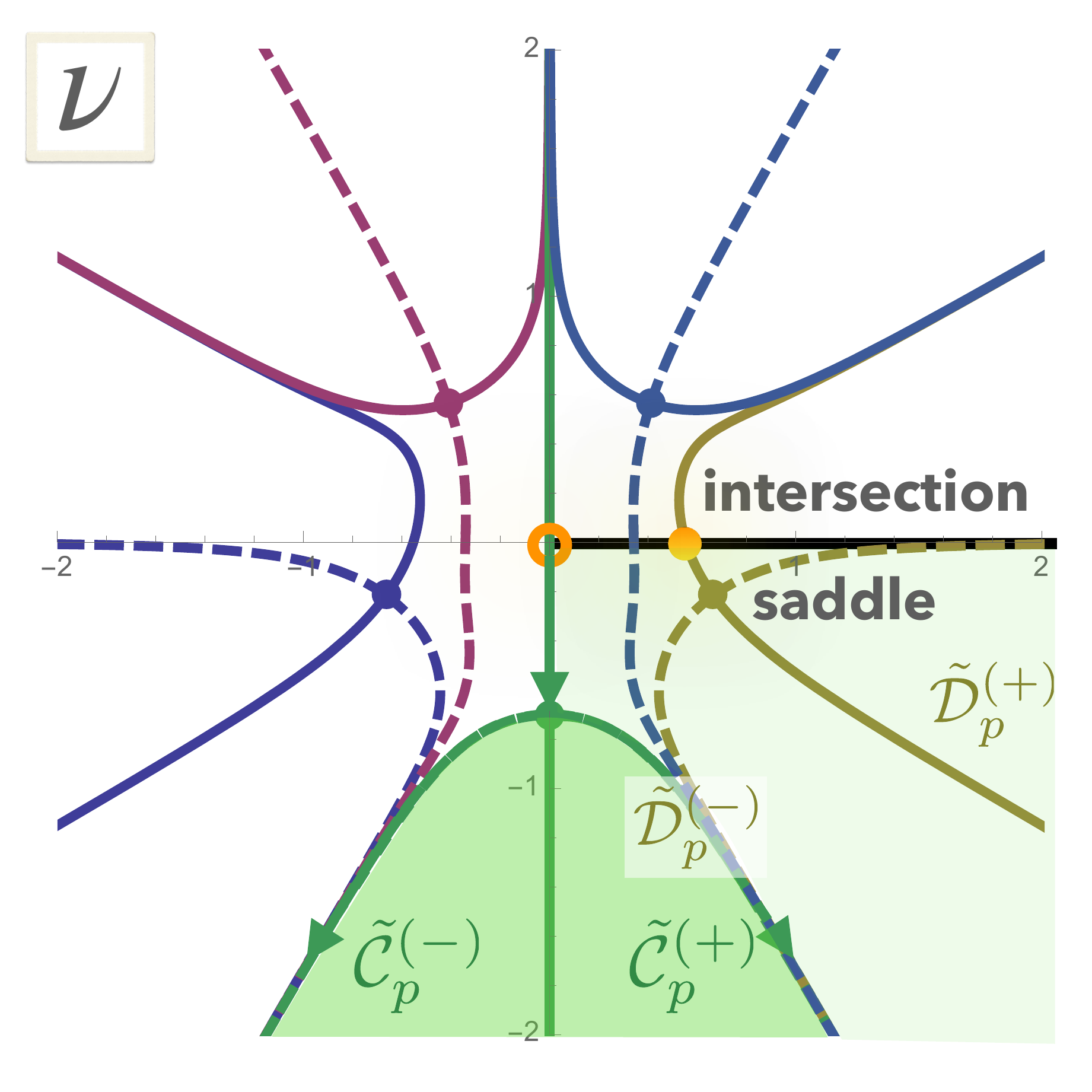}}
\\ \\
(a) $l=4$ ($g=\beta=\omega_p=1$) & &
(b) $l=6$ ($g=\beta=\omega_p=1$)
\end{tabular}
\caption{Lefschetz thimbles for $S_p(\nu)=\frac{\beta}{g}(\nu^4+i\omega_p \nu)$ ($l=4$, left panel) and for $S_p(\nu)=\frac{\beta}{g}(\nu^6+i\omega_p \nu)$ ($l=6$, right panel). For $l=4$, the integration contour $\R_+$ can be decomposed to the ascending path $\tilde{\mathcal C}_p$ from the origin and the thimble associated with the saddle point in the region $\tilde{\mathcal D}_p$ (forth quadrant). For $l=6$, the the ascending path $\tilde{\mathcal C}_p$ depends on $\arg g$. The difference $\tilde{\mathcal C}_p^+ - \tilde{\mathcal C}_p^-$ is the thimble associated with the saddle point on the negative imaginary axis, whose intersection number is $n=1$ for $\arg g < 0$ and $n=0$ for $\arg g > 0$.}
\label{fig:general_thimble}
\end{figure}

To determine the intersection numbers, 
let us consider the flow equations 
\beq
\overline{\p_s \tilde \varphi} &=& \frac{1}{g} \, \Big[ + \p_\tau + i \mu + V'(\tilde \varphi \varphi) \Big] \varphi, \label{eq:flow1_general} \\
\overline{\p_s \varphi} &=& \frac{1}{g} \, \Big[ -\p_\tau + i \mu + V'(\tilde \varphi \varphi) \Big] \tilde \varphi.\label{eq:flow2_general}
\eeq
Let us look for flows connecting the saddle points and some points on the original integration contour. 
The same argument as in the case of $V(\tilde \varphi \varphi) = (\tilde \varphi \varphi)^2$ discussed in subsection \ref{subsec:intersection}
leads to the following ansatz for the flow 
\beq
\varphi = \sqrt{\nu(s)} \, e^{\frac{2\pi p i \tau}{\beta} + i \theta} , \hs{10}
\tilde \varphi = \sqrt{\tilde \nu(s)} \, e^{-\frac{2\pi p i \tau}{\beta} - i \theta}.
\eeq
As in the previous case, we can show 
by using the conservation law for the $U(1)$ symmetry 
that
\beq
\nu(s) = \tilde \nu(s). 
\eeq
Then, the flow equation reduces to that for $\nu$
\beq
\nu'(s) = 2 |\nu(s)| \overline{\left[ \frac{V'(\nu(s)) + i \omega_p}{g}\right]}. 
\eeq
The orbit of the flow obeying this equation 
can be determined through the conservation law 
\beq
{\rm Im} \, \left[ \frac{\beta}{g} \left\{ V(\nu) + i \omega_p \nu \right\} \right] = {\rm Im} \, \mathcal S_p(\nu_{p,\sigma}) = const.
\label{eq:conseration_S_p}
\eeq
By solving this conservation law, 
we can draw a flow line from each saddle point 
on the complex $\nu$-plane
and determine the intersection number 
by checking if the flow intersects 
the positive real axis in the complex $\nu$-plane
corresponding to the original integration contour 
$\tilde \varphi = \bar \varphi$.
We can also rephrase the condition 
for the intersection number as 
\beq
n_{p,\sigma} = \left\{ 
\begin{array}{cc} 
1 & ~~~ \mbox{if $\nu_{p,\sigma} \in \tilde{\mathcal D}_p$} \\ 
0 & ~~~ \mbox{if $\nu_{p,\sigma} \not \in \tilde{\mathcal D}_p$}
\end{array} \right.,
\label{eq:intersection_nu}
\eeq
where $\tilde{\mathcal D}_p$ is the region in the $\nu$-plane 
surrounded by $\tilde{\mathcal C}_p$ and the positive real axis, 
that is, the orbit of the positive real axis under the ascending flow
(see examples in Fig.\,\ref{fig:general_thimble}). 
If the saddle point $\nu_{p,\sigma}$ is on the boundary of 
$\tilde{\mathcal D}_p$, that is, $\tilde{\mathcal C}_p$, 
the intersection number has discontinuity at 
${\rm arg} \, g = 0$. 
From the facts that 
\begin{itemize}
\item
${\rm sign} ( {\rm Im} \, \mathcal S_p(\nu_{p,\sigma}) ) = \mp 1$ for ${\rm arg} \, g = \pm 0$, ($\because \mathcal S_p(\nu_{p,\sigma}) = \frac{\beta}{g} [V(\nu_{p,\sigma}) + i \omega_p \nu_{p,\sigma}] \in \R_{\geq 0} $ at ${\rm arg} \, g = 0$ by assumption), 
\item
${\rm sign} ( {\rm Im} \, \mathcal S_p(\nu) ) = {\rm sign} ( {\rm Im} \, \mathcal S_p'(0) )$ in the neighborhood of $\tilde{\mathcal C}_p$ in $\tilde{\mathcal D}_p$,
\end{itemize}
we conclude that the discontinuity of the intersection number is given by
\beq
\Delta n_{p,\sigma} \ = \ n_{p,\sigma}^{(+)} - n_{p,\sigma}^{(-)} \ = \ 
\left\{ 
\begin{array}{cc} -
{\rm sign} ( {\rm Im} \, \mathcal S_p'(0) ) & ~~~ \mbox{if $\nu_{p,\sigma} \in \tilde{\mathcal C}_p$} \\
0 & ~~~ \mbox{if $\nu_{p,\sigma} \not \in \tilde{\mathcal C}_p$}
\end{array} \right..
\eeq
This completely cancels the discontinuity of the perturbation series \eqref{eq:ambiguity_general} 
and hence the transseries \eqref{eq:transseries_general} 
obtained through the Lefschetz thimble method
has no ambiguity. 

\subsection{Operator formalism}
So far, we have seen 
from the viewpoint of the path integral formalism
that the transseries expression 
for the generating function $Z$ takes the form in 
Eq.\,\eqref{eq:transseries_general} 
with the intersection numbers determined through 
Eq.\,\eqref{eq:conseration_S_p}. 
Here, we confirm that the transseries 
in Eq.\,\eqref{eq:transseries_general} is 
consistent with that obtained 
from the viewpoint of operator formalism.

By using the number eigenstate 
$\hat{\mathcal N} |n \rangle = n | n \rangle$, 
the generating function \eqref{eq:Z_def_general}
can be rewritten as
\beq
Z = \sum_{n=0}^\infty 
\exp \left[ - \frac{\beta}{g} V(g n) - i \beta \mu n \right].
\eeq
This expression can be further rewritten 
by using the Poisson resummation \eqref{eq:poisson} as 
\beq
Z = \frac{1}{2} + \sum_{p=-\infty}^\infty \int_0^\infty \frac{d\nu}{g} \, \exp \left( - \mathcal S_p(\nu) \right) \hs{5} \mbox{with} \hs{5} \mathcal S_p = \frac{\beta}{g} \left\{ V(\nu) + i \omega_p \nu \right\}.
\eeq
This full generating function has the same form 
as the perturbative part \eqref{eq:pert_nu} 
except for the integration contour: 
the ascending flow $\widetilde{\mathcal C}_p$ 
of $\mathcal S_p(\nu)$ for the perturbative part and
the positive real axis for the full generating function. 
By the change of integration variable 
from $\nu$ to $t=\mathcal S_p(\nu)$ given in \eqref{eq:changeofvariable_nu_t},
the generating function can be rewritten as
\beq
Z \ = \ \frac{1}{2} + \frac{1}{\beta} \sum_{p=-\infty}^\infty \int_{\mathcal C_p} dt \, \frac{e^{-t}}{V'(\nu(gt)) + i \omega_p}.
\label{eq:full_Borel}
\eeq
The integration contour $\mathcal C_p$ 
on the complex $t$-plane is the image of the positive real axis on the complex $\nu$-plane under the map 
$\nu \rightarrow t=\mathcal S_p(\nu)$. 
By deforming the integration contour $\mathcal C_p$, 
we can decompose the full generating funciton \eqref{eq:full_Borel} into 
perturbative and non-perturbative parts as
\beq
Z = \left[ \frac{1}{2} + \frac{1}{\beta} \sum_{p=-\infty}^\infty \int_0^\infty dt \, \frac{e^{-t}}{V'(\nu(gt)) + i \omega_p} \right] + 
\sum_{p,\sigma} n_{p,\sigma} \left[ \frac{1}{\beta} \int_{\mathcal C_{p,\sigma}} dt \, \frac{e^{-t}}{V'(\nu(gt)) + i \omega_p} \right], 
\label{eq:Z_operator}
\eeq
where $\mathcal C_{p,\sigma}$ is the contour surrounding 
each singularity and associated branch cut.
If $t=t_{p,\sigma}$ is a singularity, 
the corresponding intersection number is given by
\beq
n_{p,\sigma} = \left\{ 
\begin{array}{cc} 
1 & \mbox{if $t_{p,\sigma} \in \mathcal D_p$} \\ 
0 & \mbox{if $t_{p,\sigma} \not \in \mathcal D_p$} 
\end{array} \right.,
\eeq
where $\mathcal D_p$ is the region surrounded by 
the contour $\mathcal C_p$ and 
the positive real axis on the complex $t$-plane. 
This agrees with the intersection number \eqref{eq:intersection_nu}
obtained through the analysis of the flow equation 
since $\nu_{p,\sigma}$ and $\tilde{\mathcal D}_p$ 
are mapped to $t_{p,\sigma}$ and $\mathcal D_p$
under the change of variable 
$\nu \rightarrow t = \mathcal S_p(\nu)$.
Furthermore, each non-perturbative part in 
Eq.\,\eqref{eq:Z_operator} 
is related to $Z_{p,\sigma}$ in Eq.\,\eqref{eq:non_pert_all} 
by the change of variable $t = \mathcal S_p(\nu)$. 
Thus, we conclude that 
the transseries obtained 
through the Lefschetz thimble method 
\eqref{eq:transseries_general} 
is non-perturbatively complete and 
agrees with the exact result obtained in
the operator formalism \eqref{eq:Z_operator}.

\begin{figure}[!ht]
\centering
\begin{tabular}{ccc}
\fbox{
\includegraphics[width=50mm, bb = 0 0 540 540]{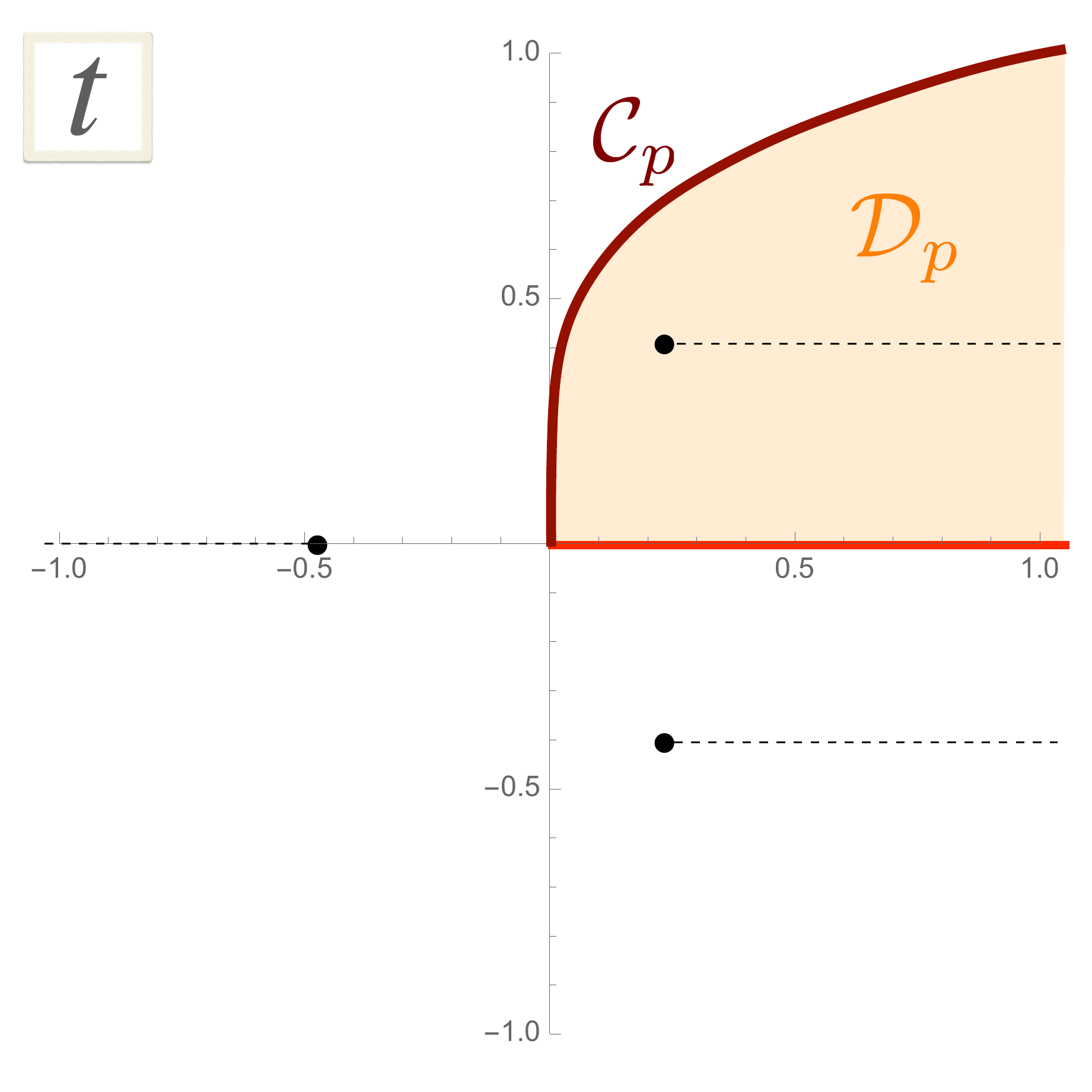}}
&~\hs{20}~&
\fbox{
\includegraphics[width=50mm, bb = 0 0 540 540]{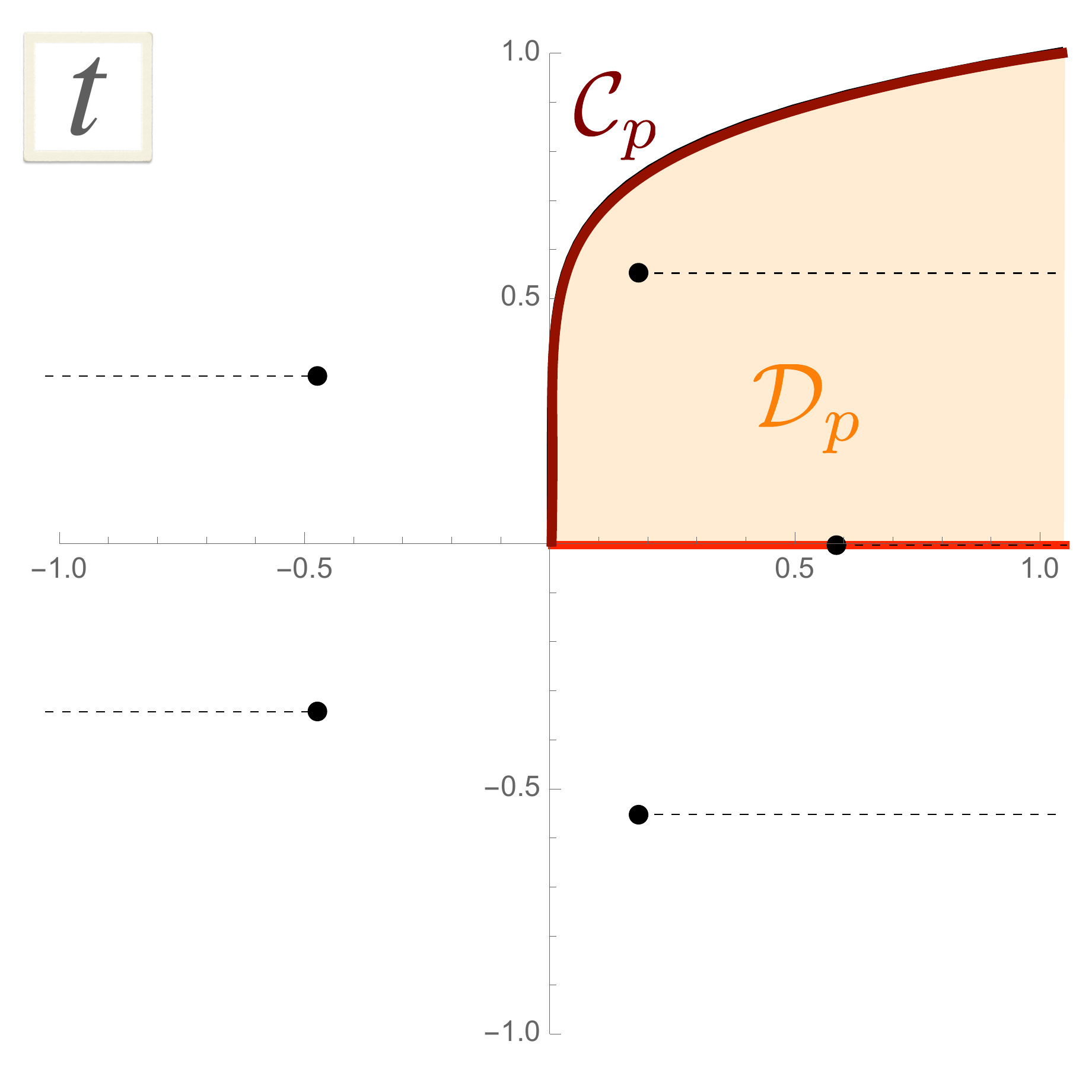}}
\\ \\
(a) $l=4$ ($g=\beta=\omega_p=1$) & &
(b) $l=6$ ($g=\beta=\omega_p=1$)
\end{tabular}
\caption{Singularities of $B_p(t)$ for $l=4$ (left panel) and $l=6$ (right panel). The integration along the positive real axis gives the perturbative part. For $l= 2~{\rm mod}~4$, there is a singularity on the positive real axis, and hence the perturabtion series is non-Borel-summable. The ambiguity is given by the integration along the branch cut emanating from the singularity on the positive real axis. The contour $\mathcal C_p$ corresponds to the full contribution. The difference between the perturbative and full contributions is given by the integration along the branch cuts emanating from the singularities contained in the region $\mathcal D_p$.}
\label{fig:general_Borel}
\end{figure}

\subsection{Example}
To illustrate the discussion in this section, 
let us consider the monomial potential as an example 
\beq
V(|\varphi|^2) = |\varphi|^{2l}, \hs{5} l \in \Z_{\geq 0}.
\eeq
In this case the generating function is given by
\beq
Z = \sum_{n=-\infty}^\infty \exp \left( - \beta g^{l-1} n^{l} - i \beta \mu n \right).
\eeq
We can easily verify that this generating function satisfies the differential equation
\beq
\left[ \frac{\p}{\p g} + (l-1) \beta g^{l-2} \left( \frac{i}{\beta} \frac{\p}{\p \mu} \right)^{l} \right] Z = 0.
\eeq
By using the power series ansatz, we can determine the perturbative part as
\beq
Z_{\rm pert} = \frac{1}{2} + \sum_{n=0}^\infty \sum_{p=-\infty}^\infty \frac{1}{i \beta \omega_p} \frac{\Gamma(l n+1)}{\Gamma(n+1)} \left( \frac{i}{\omega_p} \right)^n \left( \frac{g}{i \beta \omega_p} \right)^{n(l-1)}. 
\eeq
Since this perturbation series is factorially divergent, 
let us consider the Borel resummation 
\beq
Z_{\rm pert} &=& \frac{1}{2} + \int_0^\infty dt \, e^{-t} \sum_{n=0}^\infty \sum_{p=-\infty}^\infty \frac{1}{i \beta \omega_p} \frac{\Gamma(l n+1)}{\Gamma(n+1)\Gamma(n(l-1)+1)}  \left( \frac{i}{\omega_p} \right)^n \left( \frac{t g}{i \beta \omega_p} \right)^{n(l-1)} \notag \\
&=&  \int_0^\infty dt \, e^{-t} B(tg).
\eeq
The Borel transform $B(tg)$ takes the form of
\beq
B(tg) \ = \ \frac{1}{2} + \sum_{p=-\infty}^\infty B_p(tg),
\eeq
where $B_p(tg)$ are given by the hypergeometric functions 
\beq
B_p(tg) = \frac{1}{i \beta \omega_p} \, {}_{l-1} F_{l-2}
 \left( \left\{ \frac{1}{l}, \frac{2}{l}, \cdots, \frac{l-1}{l} \right\}, \left\{ \frac{1}{l-1}, \frac{2}{l-1}, \cdots, \frac{l-2}{l-1} \right\} , z_p \right),
\label{eq:hyper}
\eeq
with
\beq
z_p = \frac{l^l}{(l-1)^{l-1}} \frac{i}{\omega_p} \left( \frac{tg}{i \beta \omega_p} \right)^{l-1}. 
\eeq
The function $B(tg)$ becomes singular when $z_p=1$, 
i.e. it has singularities at 
\beq
t = \frac{l-1}{l} \frac{i \beta \omega_p}{g} \nu_{p,\sigma}~~~\mbox{with}~~~
\nu_{p,\sigma} = \left( \frac{-i \omega_p}{l} \right)^{\frac{1}{l-1}} \exp \left( \frac{2\pi \sigma i}{l-1} \right), \hs{5} (\sigma = 1, \cdots, l-1).
\label{eq:sing_position}
\eeq
The complex saddle points corresponding to 
this singularity on the Borel plane is given by
\beq
\varphi = \sqrt{\nu_{p,\sigma}} \, \exp \left( \frac{2\pi p i}{\beta} \tau + i \theta \right), \hs{10}
\tilde \varphi = \sqrt{\nu_{p,\sigma}} \, \exp \left( -\frac{2\pi p i}{\beta} \tau - i \theta \right).
\eeq
Note that $\nu_{p,\sigma} $ satisfies the saddle point condition  \eqref{eq:cond_sing2}
\beq
\mathcal S'(\nu_{p,\sigma})  = \frac{\beta}{g} \left[ V'(\nu_{p,\sigma}) + i \omega_p \right] = \frac{\beta}{g} \left[  l \nu_{p,\sigma}^{l-1} + i \omega_p \right] = 0. 
\eeq
We can check that the value of the action for this saddle point 
agrees with the location of the singularity \eqref{eq:sing_position}
\beq
\mathcal S_p(\nu_{p,\sigma}) = \frac{\beta}{g} \left( \nu_{p,\sigma}^l + i \omega_p \nu_{p,\sigma} \right) = \frac{l-1}{l}  \frac{i\beta \omega_p}{g} \nu_{p,\sigma}. 
\eeq
We can show that the saddle points which contribute to the generating function are those with 
\beq
\sigma = l-1 - \left \lfloor \frac{l-2}{4} \right \rfloor, \cdots, l-1. 
\label{eq:saddles_example}
\eeq
Examples of the thimble structure of $\mathcal S_p(\nu)$ 
for $l=4$ and $l=6$ 
are shown in Fig.\,\ref{fig:general_thimble}.
To see this, let us rewrite the exact generating function as
\beq
Z &=& \sum_{n=-\infty}^{\infty} \exp \left( -\beta g^{l-1} n^l - i \beta \mu n \right) \notag \\
&=& \frac{1}{2} + \frac{1}{g} \sum_{p=-\infty}^\infty \int d \nu \, \exp \left[ - \frac{\beta}{g} ( \nu^l + i \omega_p \nu ) \right]  \notag \\
&=& \frac{1}{2} + \sum_{p=-\infty}^\infty \int_{\mathcal C_p} dt \, e^{-t} B_p(tg),
\eeq
where the functions $B_p(tg)$ are the same functions as those which appeared in the Borel transform \eqref{eq:hyper}. 
The integration contours $\mathcal C_p$ are the image of the positive real axis on $\nu$-plane under the change of the variable $t = \frac{\beta}{g} (\nu^l + i \omega_p \nu)$ 
\beq
\mathcal C_p = \left\{ t \in \C \, \Big | \, {\rm Re} \, t = \frac{\beta}{g} \nu^l,~
{\rm Im} \, t = \frac{\omega_p \nu}{g} , ~\nu \in \R_{\geq 0} \right\}.
\eeq
The saddle points \eqref{eq:saddles_example}
are enclosed by the curve $\mathcal C$ 
and the positive real axis on the Borel plane,
and hence they have contributions to the generating function. 
In particular, for $ l = 2 ~{\rm mod} ~4$, the singularity with 
$q = l - 1 - (l-2)/4$ is on the positive real axis and 
hence gives rise to an ambiguity of the perturbative part. 

\section{Generalization to quantum mechanics with Integrability}
\label{sec:general}
The analysis in this paper can also be generalized 
to the multi-variable cases. 
In particular, it would be possible to obtain exact results if there exist the same number of conserved charges as degrees of freedom. 
For example, in the $N$-variable system 
described by the Lagrangian
\beq
L = \frac{1}{g} \sum_{i=1}^N \Big[ i \bar \varphi_i \p_t \varphi_i + V(|\varphi_1|^2,\cdots,|\varphi_N|^2) \Big], 
\eeq
there are $N$ conserved charges 
$\boldsymbol{\mathcal N} = (|\varphi|_1^2/g, \cdots , |\varphi|_N^2/g)$ corresponding to the phase rotations $\varphi_i \rightarrow e^{i \alpha_i}\varphi_i~(i=1,\cdots,N)$ 
and hence some exact results can be obtained. 
For example, the generating function 
\beq
Z(\boldsymbol \mu) = \tr \left[ \exp \big( -\beta \hat H - i \beta \boldsymbol \mu \cdot \hat{\boldsymbol{\mathcal N}} \big) \right], \hs{10} \left( \hat H = \frac{1}{g} V (g \hat{\mathcal N}) \right)
\eeq
satisfies the differential equation 
\beq
\left[ - \frac{1}{\beta} \frac{\p}{\p g} + \frac{1}{g^2} V \left( \hat{\boldsymbol{\nu}} \right) - \hat{\boldsymbol{\nu}} \cdot  \frac{\p}{\p \hat{\boldsymbol{\nu}}} V \left( \hat{\boldsymbol{\nu}} \right) \right] Z(\boldsymbol{\mu}) \ = \ 0, \hs{5} 
\left( \hat{\boldsymbol{\nu}} = g \frac{i}{\beta} \frac{\p}{\p \boldsymbol{\mu}} \right),
\eeq
where $\boldsymbol \mu = (\mu_1,\cdots,\mu_N)$ 
are chemical potentials for the conserved charges. 
The perturbation series can be determined from this differential equation with the initial condition 
$Z_{g=0} = \prod_{i=1}^N (1-e^{-i \beta \mu_i})^{-1}$. 
In general, the Borel transform of the perturbation series 
has singularities corresponding to non-perturbative saddle points.  Such saddle point solutions of the Wick rotated equation of motion can be obtained by using the conservation laws
\beq
\varphi_i = \sqrt{\nu_i} \, \exp \left( i \boldsymbol{\omega}_{\mathbf p} \tau + i \boldsymbol{\theta} \right)_i , \hs{10}
\tilde \varphi_i = \sqrt{\nu_i} \, \exp \left( -i \boldsymbol{\omega}_{\mathbf p} \tau - i \boldsymbol{\theta} \right)_i,
\eeq
where $\boldsymbol \theta = (\theta_1,\cdots,\theta_N)$ are moduli parameters (integration constants) and we have defined 
\beq
\omega_{\mathbf p} = \boldsymbol{\mu} + \frac{2\pi}{\beta} \mathbf p, \hs{19} \mathbf p = (p_1,\cdots, p_N) \in \Z^N. 
\eeq
The values of $\boldsymbol{\nu} = (\nu_1,\cdots,\nu_N)$ 
are determined from the conditions
\beq
\nu_i \frac{\p}{\p \nu_i} \mathcal S_{\mathbf p}(\boldsymbol \nu) = 0 \hs{10} (i=1,\cdots,N),
\label{eq:saddle_cond_multi}
\eeq
where $\mathcal S_p(\boldsymbol{\nu})$ is the $N$-variable function 
\beq
\mathcal S_p(\boldsymbol{\nu}) = \beta V(\boldsymbol{\nu}) + i \beta \boldsymbol{\omega}_{\mathbf p} \cdot \boldsymbol{\nu}.  
\eeq
The contribution from these saddle points can also 
be determined from the one-loop determinant 
by solving the differential equation. 
The intersection number can be determined by the flow equation. 
Using the ansatz
\beq
\varphi_i = \sqrt{\nu_i(s)} \, \exp \left( i \boldsymbol{\omega}_{\mathbf p} \tau + i \boldsymbol{\theta} \right)_i , \hs{10}
\tilde \varphi_i = \sqrt{\nu_i(s)} \, \exp \left( -i \boldsymbol{\omega}_{\mathbf p} \tau - i \boldsymbol{\theta} \right)_i,
\eeq
we can reduce the flow equation for $(\varphi,\tilde \varphi)$ to that for $\nu_i$ 
\beq
\frac{1}{2|\nu_i|} \overline{\frac{\p \nu_i}{\p s}} = \frac{1}{\beta} \frac{\p \mathcal S_p}{\p \nu_i}. 
\eeq
Combining the saddle point contributions and the intersection numbers, we can construct the transseries for the generating function. 

In the operator formalism, the generating function is given by
\beq
Z = \sum_{n_1=0}^\infty \cdots \sum_{n_N=0}^\infty \exp \left[ -\frac{\beta}{g} V( g n_1,\cdots,gn_N) - i \beta \mu_i n_i \right].
\eeq
By applying the Poisson resummation formula \eqref{eq:poisson} to each summation, the generating function can be rewritten as
\beq
Z = \mathcal Z_0 + \mathcal Z_1 + \cdots + \mathcal Z_N, 
\eeq
where $\mathcal Z_m$ is given by integrals over $m$-face of the region $\nu_i \geq 0~(i=1,\cdots,N)$ 
\beq
\mathcal Z_m = \frac{1}{g^m 2^{N-m}} \sum_{p_1 \in \Z} \cdots \sum_{p_m \in \Z} \int_0^\infty d\nu_1 \cdots d\nu_m \, \exp \left( - \mathcal S_p \right)_{\nu_{m+1}=\cdots=\nu_N=0} + \{\mbox{permutations}\}.
\label{eq:Z_multi}
\eeq
Applying the Lefschetz thimble method to each integral, 
we can confirm the correspondence between the path integral and operator formalisms. 
For example, each solution of the saddle point conditions \eqref{eq:saddle_cond_multi} in the path integral formalism is a saddle point of one of $\mathcal Z_m$. 
In this way, we can check the correspondence of saddle points, gradient flows, and turning points in the path integral and operator formalisms. 
Note that, in general, 
the saddle point configurations satisfying \eqref{eq:saddle_cond_multi}
are complex saddle points.
This shows that the complexification of the path integral
is indispensable for obtaining exact transseries. 

It would also be possible to generalize the discussions 
to general integrable systems,
where the action can be rewritten
by using the action-angle variables $(\boldsymbol{\nu}, \boldsymbol{\vartheta})$ as
\beq
S_E = \int d\tau \Big[ i \boldsymbol \nu \cdot \partial_\tau \boldsymbol{\vartheta} - H(\boldsymbol{\nu}) - i \boldsymbol{\mu} \cdot \boldsymbol{\nu} \Big].
\label{eq:action_able}
\eeq
Assuming that $\boldsymbol \vartheta$ is on the invariant torus
$\boldsymbol{\vartheta} = \frac{2 \pi \tau}{\beta} \mathbf p + \boldsymbol{\theta}~(\mathbf p \in \Z^N)$, 
the saddle point condition and flow equation for $S_{\rm E}$ reduces to those for the function 
\beq
\mathcal S_p = \beta H + i \beta \boldsymbol{
\omega}_{\mathbf p} \cdot \boldsymbol{\nu}, \hs{10}
\left( \boldsymbol{\omega}_{\mathbf p} = \boldsymbol{\mu} + \frac{2 \pi \tau}{\beta} \mathbf p \right).  
\eeq
On the other hand, in the operator formalism, 
the generating function can be rewritten into a form
similar to \eqref{eq:Z_multi} depending on 
the details of the quantization conditions 
of the conserved charges $\mathbf \nu$. 
Then, applying the Lefschetz thimble method, 
we can confirm the correspondence 
between the path integral and operator formalisms. 
In this way, it would be possible to show that 
the Lefschetz thimble formalism gives exact results 
which are consistent with the operator formalism
in general integrable systems. 

\section{Conclusions and discussion} 
\label{sec:conclusion}
In this paper, we have discussed the resurgence structure 
of the generating function for the conserved charge 
in the $U(1)$ symmetric first-order time derivative systems. 
We have explicitly evaluated the path integral 
for the generating function 
by following the Lefschetz thimble method 
with the help of the differential equation 
which enables us to determine the all-order perturbation series around each saddle point. 
We have checked that the results obtained through the Lefschetz thimble method were consistent with the exact expressions obtained in the operator formalism. 
This fact indicates the non-perturbative completeness of the Lefschetz thimble method. 

We have seen that the resurgence structure of 
the quantum mechanical system considered in this paper 
can be correctly captured by the Lefschetz thimble method.
It would be interesting to generalize the discussion to 
the more general quantum mechanical systems with explicit analytic solutions. 
The key point that enables us to analyze exact results explicitly is
integrability, i.e., the property that the number of degrees of freedom is 
the same as that of conserved charges. 
It would be possible to generalize our discussion to 
general integrable quantum mechanical systems. 
The explicit analysis of thimbles of the action written in terms of the action-angle variables \eqref{eq:action_able} is important future work. Quantum mechanics with a single degree of freedom is one of the simplest classes of models where the action can be rewritten into the form \eqref{eq:action_able} by using the conserved energy. 
Therefore, we can apply the analysis in this paper to such systems. It would be interesting to analyze the relationship between the method discussed in this paper and the exact WKB analysis. 

It is also important to generalize the thimble analysis 
to the integrable quantum field theories. 
The non-linear Schr\"odinger system in two dimensions, whose 1d reduction is the model discussed in Sec.\,\ref{sec:4th}, is one of the examples of integrable field theories. 
It is more non-trivial to correctly determine the resurgence structure of field theories due to the existence of so-called renormalons 
\cite{Marino:2019eym}, 
whose relation to saddle point configurations has not yet been well understood. 
Understanding the resurgence structure, in particular, the renormalons in the path integral formalism of exactly solvable models is important future work. 

\begin{acknowledgements}
This work is supported by the Ministry of Education, Culture, 
Sports, Science, and Technology(MEXT)-Supported Program for the 
Strategic Research Foundation at Private Universities ``Topological 
Science" (Grant No. S1511006) 
and 
by the Japan Society for the Promotion of Science (JSPS) 
Grant-in-Aid for Scientific Research (KAKENHI) Grant Number 
(18H01217).
This work is also supported in part by JSPS KAKENHI Grant Numbers 
JP18K03627, JP20F20787, JP21K03558 (T.\ F.), JP19K03817 (T.\ M.) and JP22H01221 (M.\ N.).
S.\ K. is supported by the Polish National Science Centre grant
2018/29/B/ST2/02457.
\end{acknowledgements}

\appendix

\section{Lefschetz thimble method}
\label{appendix:Lefschetz}
In this appendix, we recapitulate the Lefschetz thimble method. 
Suppose that we are interested in a path integral of the form of 
\beq
Z = \int_{\mathcal C} \D \phi \, \exp \left( - S[\phi] \right).
\eeq
By deforming the integration contour
$\mathcal C = \{ \phi \in \R \}$, 
this path integral can be decomposed as
\beq
Z = \sum_{\sigma \in \mathfrak S} n_\sigma Z_\sigma, 
\eeq
where $\mathfrak S$ denotes the set of all 
the saddle points of $S[\phi]$, 
that is, the solutions of the complexified equation of motion
\beq
\frac{\delta S}{\delta \phi} = 0. 
\eeq
The contribution associated with each saddle point 
is given by the path integral over 
the Lefschetz thimble $\mathcal J_\sigma$: 
\beq
Z_\sigma \equiv \int_{\mathcal J_\sigma} \D \varphi \, \exp (-S).
\label{eq:Z_sigma}
\eeq
The thimble $\mathcal J_\sigma$ associated 
with the saddle point $\phi_{\sigma}$ is the set of points 
in the complexified configuration space 
which can be reached from the saddle point by the flow 
\beq
\frac{d \phi}{ds} = \overline{\frac{\delta S}{\delta \phi}}, \hs{10}
\lim_{s \rightarrow -\infty} \phi = \phi_{\sigma}, 
\label{eq:flow}
\eeq
where $s$ is a formal flow parameter. 
Note that ${\rm Re} \, S$ is strictly increasing and ${\rm Im} \, S$ is constant along the upward flow 
\beq
\frac{d}{d s} {\rm Re} \, S > 0 , \hs{10} 
\frac{d}{d s} {\rm Im} \, S = 0. 
\eeq
The coefficients $n_\sigma$ indicate
how the original integration contour $\mathcal C$ is decomposed: 
\beq
\mathcal C = \sum_{\sigma \in \mathfrak S} n_\sigma \mathcal J_\sigma. 
\eeq
They can also be defined as the intersection numbers between $\mathcal C_{\R}$ 
and ``the dual thimble $\mathcal K_\sigma$" defined as the set of points 
which flows to the saddle point $\sigma$: 
\beq
\frac{d \varphi}{dt} = \overline{\frac{\delta S}{\delta \varphi}}, \hs{10}
\lim_{t \rightarrow \infty} \varphi = \varphi_{\rm sol,\,\sigma}.
\eeq
Since the thimble $\mathcal J_\sigma$ and its dual $\mathcal K_\sigma$ are 
defined in terms of the flow, 
it follows that the real and imaginary parts of the complexified action satisfy
\beq
{\rm Re} \, S |_{\mathcal J_\sigma} \geq {\rm Re} \, S |_{\rm sol, \sigma} \geq {\rm Re} \, S |_{\mathcal K_\sigma}, \hs{10}
{\rm Im} \, S |_{\mathcal J_\sigma} = {\rm Im} \, S |_{\rm sol, \sigma} = {\rm Im} \, S |_{\mathcal K_\sigma}.
\eeq 
These properties imply that $\mathcal J_\sigma$ and $\mathcal K_\sigma$ 
intersect exactly once at the saddle point $\sigma$, 
and $\mathcal J_\sigma$ cannot intersect with $\mathcal K_{\sigma'}~(\sigma' \not = \sigma)$ 
since ${\rm Im} \, S |_{\mathcal J_\sigma} \not = {\rm Im} \, S |_{\mathcal K_{\sigma'}}$ for a generic action. 
Therefore, the intersection pairing of $\mathcal J_\sigma$ and $\mathcal K_{\sigma'}$,
regarded as middle dimensional relative homology cycles, is given by
\beq
\langle \mathcal J_\sigma, \mathcal K_{\sigma'} \rangle = \delta_{\sigma \sigma'}. 
\eeq
Using this pairing, we can calculate the coefficients $n_\sigma$ as the intersection number 
of the original contour $\mathcal C_\R$ and the dual thimble $\mathcal K_\sigma$:
\beq
n_\sigma = \langle \mathcal C_\R , \mathcal K_\sigma \rangle.
\eeq

The perturbative part of the partition function corresponds to 
$Z_0$ defined as the path integral over the thimble $\mathcal J_0$
emanating from the trivial vacuum configuration. 
Non-perturbative contributions are given by the path integral over 
thimbles associated with non-trivial saddle points $\sigma$. 
It is often the case that the partition function 
for a real positive coupling constant $g$ is on the Stokes line, 
i.e., the line on which the thimbles $\mathcal J_\sigma$ 
and the coefficients $n_\sigma$ change discontinuously 
when we vary the coupling constant in the complex $g$ plane. 
If $\mathcal J_0$ jumps on the real axis (${\rm Im} \, g = 0$), 
the perturbative part $Z_0$ has an ambiguity depending on 
how we take the limit ${\rm Im} \, g \rightarrow \pm 0$. 
However, the original partition function $Z$ has no ambiguity 
since it is defined independently of $\mathcal J_\sigma$ and $n_\sigma$. 
Therefore, the ambiguity of $Z_0$ has to be canceled by 
those associated with other non-trivial saddle points. 
In the case of $\C P^1$ quantum mechanics, such saddle points correspond to the bion configurations \cite{Dunne:2012ae, Dunne:2012zk, Misumi:2014jua, Misumi:2014bsa}, and their contributions have ambiguities as can also be seen in the result of the Gaussian approximation \eqref{eq:one-loop}. 
We will see below that the ambiguity of the bion contribution 
originates from the discontinuous change of the intersection number $n_\sigma$ associated with the bion saddle points.

\section{Differential equation for generating function}
\label{appendix:diff_eq}
Here we determine the higher-order correction around the non-perturbative saddle point in the case of the quartic potential. The leading order contribution from the $p$-th saddle point is given by
\beq
Z_p = \sqrt{\frac{2\pi}{\beta g}} \exp \left( - \frac{\beta}{2g} \omega_p^2 \right) \Big[ 1 + \mathcal O(g) \Big]. 
\label{eq:one-loop}
\eeq
To determine the higher order corrections, 
let us solve the differential equation \eqref{eq:diff_eq}
by assuming the power series ansatz
\beq
Z_p = \sqrt{\frac{2\pi}{\beta g}} \exp \left( - \frac{\beta}{2g} \omega_p^2 \right) \Big[ 1 + a_1 g + a_2 g^2 + \cdots \Big], 
\eeq
where the coefficients $a_n~(n=1,2,\cdots)
$ are functions of $\mu$.
Substituting into the differential equation \eqref{eq:diff_eq},
we obtain the following recursive differential equation
\beq
\left( \omega_p \p_\mu + n \right) a_n = \frac{1}{2\beta} \p_\mu^2 a_{n-1}. 
\eeq 
The general solution is given by
\beq
a_n = \sum_{m=1}^n \frac{c_m}{\omega_p^m} \left( -\frac{1}{2\beta \omega_p^2} \right)^{n-m} \frac{(2n-m-1)!}{(n-m)!(m-1)!},
\eeq
where $c_m$ are arbitrary constants. 
We can show from the path integral expression 
in the quartic potential model 
that the coefficients $a_n$ are non-singular at $\omega_p = 0$. 
Thus, we conclude that $c_n=0$ and there is no correction to the leading order contribution \eqref{eq:one-loop}, 
that is, the non-perturbative contributions are one-loop exact. 

In general, the non-perturbative contributions of 
non-trivial saddle points take the form of
\beq
Z_{p,\sigma} = \sqrt{\frac{2\pi}{\beta g V''(\nu_{p,\sigma})}} \exp \left( - \frac{\mathcal S_p(\nu_{p,\sigma})}{g} \right) \Big[ 1 + a_1 g + a_2 g^2 + \cdots \Big]. 
\eeq
The coefficients $a_n$ can be determined by the differential equation \eqref{eq:diff_eq_general}, 
which reduces to the recursive differential equation of the form of
\beq
(X \p_\mu + n) a_n = Y_n(\mu),
\label{eq:diff_a_n}
\eeq
where $Y$ is a function of $\mu$ determined once the solutions $a_i~(i=1,\cdots,n-1)$ are given and 
$X(\omega_p)$ is a function of $\omega_p$ such that 
\beq
\lim_{\omega_p \rightarrow 0} X(\omega_p) = 0.
\eeq
This implies that the general solution of the homogeneous equation $(X \p_\mu + n) a_n = 0$ is singular in the limit $\omega_p \rightarrow 0$. Therefore, we can uniquely fix the solution of the differential equation \eqref{eq:diff_a_n} by requiring that it is regular in the limit $\omega_p \rightarrow 0$.
Thus, all the coefficients around the saddle points can be uniquely determined by solving the differential equation \eqref{eq:diff_eq_general}.

\bibliographystyle{ieeetr}
\bibliography{1d.bib}

\end{document}